\documentclass[preprint,aps,prd,showpacs,nofootinbib,tightenlines]{revtex4-1}
\usepackage{amsmath}
\usepackage{amssymb}
\usepackage{epsfig}
\usepackage{color,graphicx}
\usepackage{bm}
\usepackage{times}
\usepackage{braket}
\usepackage{slashed}
\usepackage{hyperref}
\usepackage{float}

\newcommand{\beq}{\begin{eqnarray}}
\newcommand{\eeq}{\end{eqnarray}}
\newcommand{\non}{\nonumber\\ }

\newcommand{\jpsi}{J/\psi}

\newcommand{\etap}{\eta^{(\prime)}}
\newcommand{\psl}{ p \hspace{-2.0truemm}/ }

\newcommand{\epsl}{ \epsilon \hspace{-2.0truemm}/ }

\def \cpc{ {\bf Chin. Phys. C } }

\def \epjc{{\bf Eur. Phys. J. C} }
\def \jhep{{\bf JHEP } }
\def \jpg{ {\bf J. Phys. G} }

\def \npb{ {\bf Nucl. Phys. B} }
\def \plb{ {\bf Phys. Lett. B} }

\def \prd{ {\bf Phys. Rev. D} }
\def \prl{ {\bf Phys. Rev. Lett.}  }

\def \rmp{ {\bf Rev. Mod. Phys. } }
\def \SB{ {\bf Sci. Bull. } }
\def \csb{ {\bf Chin. Sci. Bull. } }
\def \epjwc{ {\bf EPJ Web Conf. } }
\def \jhep{ {\bf JHEP } }

\definecolor{Blue}{rgb}{0.,0.,1.}

\definecolor{nicegreen}{rgb}{0.1,0.5,0.1}

\hypersetup{colorlinks,citecolor=nicegreen,linkcolor=nicegreen}

\begin{document}


\title{Semileptonic decays $B/B_s \to (D^{(*)},D_s^{(*)}) l \nu_l$ in the PQCD approach with the lattice QCD input}
\author{Xue-Qing Hu$^{1}$ } \email{hu-xueqing@qq.com}
\author{Su-Ping  Jin$^{1}$ }  \email{2223919088@qq.com}
\author{Zhen-Jun Xiao$^{1,2}$  } \email{xiaozhenjun@njnu.edu.cn}
	\affiliation{1.  Department of Physics and Institute of Theoretical Physics,
		Nanjing Normal University, Nanjing, Jiangsu 210023, People's Republic of China,}
	\affiliation{2. Jiangsu Key Laboratory for Numerical Simulation of Large Scale Complex Systems,
	Nanjing Normal University, Nanjing 210023, People's Republic of China}
	\date{\today}
\begin{abstract}
In this paper, we studied the semileptonic $B/B_s \to (D^{(*)},D_s^{(*)}) l\nu_l$  decays in the framework of the standard model (SM) ,   by employing
the perturbative QCD (PQCD) factorization formalism combining with the lattice QCD inputs of the relevant transition form factors.
We calculated the branching ratios ${\cal B}(B_{(s)}  \to D_{(s)}^{(*)} l \nu_l )$ with $l=(e,\mu.\tau)$ ,
the ratios of the branching ratios $R(D^{(*)})$ and $R(D_s^{(*)} )$ and other physical observables $P_\tau(D_{(s)}^{(*)})$, $F_L(D^*_{(s)})$ and $A_{FB}(\tau)$.
The  ``PQCD+Lattice"  predictions for ${\cal B}(B \to D^{(*)} l\nu_l)$  and $R(D^{(*)})$  do  agree well with those currently available
experimental measurements within the errors.
For the ratios $R(D_s)$ and $R(D_s^*)$,  the  "PQCD+Lattice"  predictions  agree well with other  known predictions.
For both  $P_\tau(D^*)$ and $F_L(D^*)$,  our  theoretical predictions agree well with the measured values within errors.
Our theoretical  predictions  about the considered semileptonic  $B/B_s$ decays   could  be tested in  the near future LHCb and the Belle II experiments.
\end{abstract}


\maketitle


\section{Introduction}\label{sec:1}

The studies for the semileptonic decays of $B/B_s$ meson  do play an essential role in testing
the standard model (SM) as well as in searching for new physics (NP) beyond the SM,
since the lepton flavor universality (LFU) can be examined though this kind of decay modes.
The LFU means that all electroweak gauge bosons  ( $Z^0$, $\gamma$ and $W^\pm$)
have equivalent couplings to three generation leptons, and the only difference arises due to the mass differences: $m_e <  m_\mu  \ll   m_\tau$.
Therefore, if the experiments could find   some signals for the lepton flavor violation (LFV), it would be a true challenge to the SM.

The first observation of the $R(D)$ and $R(D^*)$ anomalies for the semileptonic decays $B\to D^{(*)}  l \nu_l $ with $l^-=(e^-,\mu^-,\tau^-)$
in 2012  by BaBar Collaboration \cite{babar2012,babar2013}
has invoked intensive studies for $B \to D^{(*)} l \nu_l $ decays in the framework of the SM \cite{sm4,sm1,sm2,sm3,sm5,sm6,pqcd1,pqcd3}
and various new NP models\cite{np1,np2,np3,np4,np5,np6,np7,np8}.
When more measurements are reported by both Belle and LHCb Collaboration
\cite{belle2015,belle2017,belle2018,belle2019a,belle2019b,lhcb2017,lhcb2018,belle2019a,belle2019b},
however,  the deviations between  the measured  $R(D)$ and $R(D^*)$ and  the SM predictions become a little narrow
now \cite{Caria1903,Hicheur1908,hfag2017,hflav2019}:
\begin{enumerate}
\item[(1)]
The latest Belle measurements \cite{belle2019b}  alone  exhibit   an excellent consistency  with the averaged SM predictions \cite{sm4,sm1,sm2,hflav2019}:
\beq
R(D) &=&  \left \{ \begin{array}{ll}
 0.307\pm  0.037(stat.)\pm 0.016(syst.),  &  {\rm Belle\ \  [24]}\\
0.299 \pm 0.003,   & {\rm SM\ \  [29]}\\    \end{array} \right.  , \label{eq:rd01}  \\
R(D^*) &=&  \left \{ \begin{array}{ll}
 0.283\pm  0.018(stat.)\pm 0.014(syst.),  &  {\rm Belle \ \  [24]},  \\
0.258 \pm 0.005,   & {\rm SM\ \ [29]}\\    \end{array} \right.  , \label{eq:rds01}
\eeq
they are compatible  within one standard deviation \cite{belle2019b,hflav2019}.

\item[(2)]
The combined analysis of currently available measured $R(D)$ and $R(D^*)$ with the inclusion of the new Belle results \cite{belle2019b} gives the following world
averaged values:
\beq
R(D)^{\rm Exp}=0.340\pm 0.027\pm 0.013, \qquad R(D^*)^{\rm Exp}=0.295\pm 0.011\pm 0.008, \label{eq:rdexp01}
\eeq
the discrepancy decreases from the previous $3.8\sigma$ to $3.1\sigma$ with respect to the SM expectations \cite{Caria1903,Hicheur1908,hflav2019}.

\item[(3)]
Besides  the ratios $R(D^{(*)})$,  some relevant physical  observables, such as   the longitudinal polarization of the tau lepton $P_\tau(D^*)$,
the fraction of $D^*$ longitudinal polarization  $F_L(D^*)$,  have been measured very recently by Belle Collaboration \cite{prl118-801,prd97-012004,1903tau}:
\beq
P_\tau( D^*) &=& -0.38 \pm 0.51(stat.) ^{+0.21}_{-0.16}(syst.),  {\rm [30,31]}, \label{eq:ptaustar} \\
F_L(D^*) &=&  0.60 \pm 0.08(stat.) \pm  0.04 (syst.) , {\rm [32]}.
\label{eq:flstar}
\eeq
They are compatible  with the SM predictions $P_\tau( D^*) = -0.497 \pm 0.013$ \cite{ptau2},
$F_L(D^*) =  0.441(6)$ \cite{huang18} and  $ 0.457(10)$ \cite{bhatt18} for $B \to D^* \tau  \bar{\nu}_\tau$ decays.

\end{enumerate}
It is evident to see from above points that  although the anomalies about $R(D^{(*)})$ become less serious recently,  it is still a
sizeable discrepancy with the SM expectations, and it must be  investigated with
complementary and more precise measurements in order to make a conclusion about it.
In addition to the $B \to D^{(*)} l \nu_l$ decays,   the $B_s \to D_s^{(*)} l \nu_l$ decay mode  is  one of the best choices for crossing examination.
The systematic theoretical studies  and  experimental measurements for the semileptonic decays of $B_s$ meson are therefore very important
and worth of  doing immediately.

As illustrated by the Feynman diagrams in Fig.~\ref{fig:fig1}, the $B_s \to D_s^{(*)} l^+ \nu_l$ decays are closely related with those
$B \to D^{(*)} l^+ \nu_l$ decays through the SU(3) flavor symmetry. They are all controlled by the same $b \to c l \nu$ transitions at
the quark level, but with a different spectator quark: $(u,d)$ or $s$ quark.
In the limit of SU(3) flavor symmetry,   consequently, these two sets of semileptonic decays must have very similar properties.
If current anomalies in $B\to D^{(*)} l \nu_l$ decays are indeed induced by the new physics contributions,  it must show up in the
$B_s \to D_s^{(*)} l \nu_l  $ decays.
It is  therefore necessary and very interesting to study  $B_s \to D^{(*)}_{s} l \nu_l$ decays and to measure the corresponding ratios  $R(D_s^{(*)})$ and other relevant physical
observables such as  $P_\tau(D_s^{(*)})$, $F_L(D_s^*)$ and $A_{FB}(\tau)$,  in order to check if there exist any similar deviations.

In the framework of the SM,  as is well known,  the central issue for the calculations of such semileptonic decays  is the estimation for  the values and shapes of the
relevant form factors $(F_{+,0}(q^2), V(q^2),A_{0,1,2}(q^2))$ for the $B_{(s)}  \to D_{(s)}^{(*)}$ transitions.
However,   the calculations for these form factors are not an easy task and can not be estimated reliably in  the whole range of  the  momentum
$q^2$ carried by the lepton pair by using one method,  the extrapolation is indispensable.
There are many traditional methods or approaches to estimate the relevant form factors and then provide their own predictions for
the ratios $R_{D^{(*)}}$, for examples, the heavy quark effective theory (HQET)\cite{sm1,sm2},
the light cone sum rules (LCSR)\cite{lcsr1,lcsr2,lcsr3}, the lattice QCD (LQCD)\cite{lattice1,lattice2,lattice3,lattice4,lattice5} and the perturbative
QCD factorization approach (PQCD) \cite{pqcd1,pqcd3}.

In recent years, the PQCD factorization approach has been used to study the various kinds of  the semileptonic decays of $B/B_s/B_c$ mesons
for example as being done in  Refs.~\cite{pqcd1,pqcd3,pqcd2,Wang12a,Wang13a,Xiao14a,Wang13b,Wang14a,Hu19a}.
However, just like many other theoretical approaches, the form factors evaluated by using the PQCD approach are only reliable at the low $q^2$ region.
Therefore, the extrapolations must be done in order to cover the whole range of  $q^2$ of the form factors $f_i(q^2)$.
In order to improve the reliability of the size and shape of all the six relevant form
factors obtained by employing the PQCD approach,  we here will include the lattice QCD inputs at the end point $q^2_{max}$
so that the extrapolation from the PQCD predictions at the low $q^2$ region to the high $q^2$ region  become reliable too.

In Refs.~\cite{pqcd1,pqcd3},  the $B \to D^{(*)} l \nu_l$ decays have been studied by employing the PQCD approach only \cite{pqcd1}
or with  the inclusion of the  lattice QCD input \cite{pqcd3}.
In Ref.~\cite{pqcd2},  the $\bar{B}^0_s \to D_s^{(*)} l^- \bar{\nu}_l$ decays have been studied by employing the PQCD approach
without  the lattice QCD input.  In this paper,  we will study the semileptonic decays of $B$ and $B_s$ mesons simultaneously and focus on the
following  three tasks:
\begin{enumerate}
\item[(1)]
For both  $B_s \to D_s^{(*)}$ and $B\to D^{(*)}$ and transitions,   we evaluate the relevant form factors  in the  low $0\leq q^2\leq m_l^2$ region
by using the PQCD factorization approach, and then  include the lattice QCD inputs at the end point $q^2_{max}$ in the fitting  process
and extrapolate those form factors to the entire momentum region by employing the Bourrely-Caprini-Lellouch (BCL) parametrization
\cite{bcl09,jhep1905-094} instead of the pole models used previously in Refs.~\cite{pqcd1,pqcd2,pqcd3}.

\item[(2)]
In addition to the calculation for the branching ratios and  the ratios $R(D_s^{(*)})$ and  $R(D^{(*)})$, we will also
calculate other three kinds of physical observables ( they are not considered in previous works \cite{pqcd1,pqcd2,pqcd3} ) for the
decays of  $B$ and $B_s$ mesons:
the longitudinal polarization of the tau lepton $P_\tau(D_{(s)}^{(*)})$,   the fraction of $D_{(s)}^*$ longitudinal polarization  $F_L(D_{(s)}^*)$
and the forward-backward asymmetry of the tau lepton $A_{FB}(\tau)$ in both the ordinary PQCD approach and  the ``PQCD +Lattice''
approach (i.e. the PQCD approach with the inclusion of the lattice QCD input for form factors).

\item[(3)]
We will present our theoretical predictions, compare them with those currently available experimental measurements or the theoretical predictions
obtained by using other different theories or models.

\end{enumerate}

This paper is organized as follows: In section \ref{sec:2}, we briefly review the kinematics of the $B^0_s \to D_s^{(*)} l^+ \nu_l$.
The calculations of the form factors for the $B^0_s \to D_s^{(*)}$ transitions are given then. We use the Lattice QCD results at $q^2_{max}$
given by the HPQCD group\cite{lattice5} as the inputs  in our extrapolation.
The explicit expressions of the differential decay rates and additional physical observables are also given in Sec.~\ref{sec:2}.
In section \ref{sec:3}, we present the theoretical predictions for all considered physical observables obtained by using the PQCD approach,
the ``PQCD+Lattice" approach and some typical different models. A short summary will be given in the final section.
\section{Theoretical framework}\label{sec:2}

\subsection{Kinematics and the wave functions}
\begin{figure}[htbp]
  \centering
  \vspace{-2cm}
  \includegraphics[width=16cm]{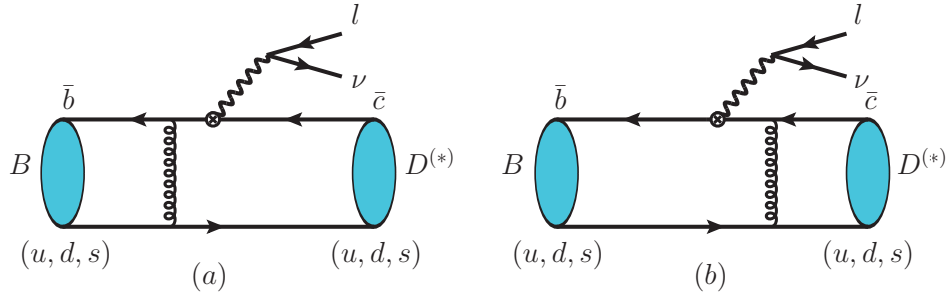}\\
  \vspace{-17cm}
  \caption{(Color online) In PQCD approach, the leading order Feynman diagrams for the semileptonic decays $B_{(s)} \to D_{(s)} l^+ \nu_l $
  with $l=(e,\mu,\tau)$.}
  \label{fig:fig1}
\end{figure}

In the PQCD approach, the tree-level Feynman diagrams for $B_{(s)} \to D^{(*)}_{(s)} l \nu$ decays
\footnote{Throughout this paper the symbol $B_{(s)}$ describes  both  $B=(B_u,B_d)$  and $B_s$ mesons.} are
shown in Fig~\ref{fig:fig1}. We define the $B_{(s)}$ meson momentum as $p_1$, the $D_{(s)}/D^*_{(s)}$ meson
momentum as $p_2$, and the polarization vectors $\epsilon_{L,T}$ of the $D_{(s)}^*$ at the $B_{(s)}$ meson rest frame
as in Ref.~\cite{Li}.
\beq
p_1&=&\frac{m_{B_{(s)}}}{\sqrt{2}}(1,1,0_\bot),\quad p_2=\frac{r \cdot m_{B_{(s)}}}{\sqrt{2}} (\eta^+,\eta^-,0_\bot),\non
\epsilon_L&=&\frac{1}{\sqrt{2}}(\eta^+,-\eta^-,0_\bot), \quad \epsilon_T=(0,0,1).
\label{eq1}
\eeq
while $\epsilon_L$ and $\epsilon_T$ denotes the longitudinal and transverse polarization of the $(D^*, D_s^*)$ mesons, respectively.
The parameter $\eta^{\pm}$ and $r$ are defined as:
\beq
\eta^{\pm}=\eta \pm \sqrt{\eta^2-1},\quad \eta=\frac{1}{2r}(1+r^2-\frac{q^2}{m_{B_{(s)}}^2}),\quad r=\frac{m_{D_{(s)}^{(*)}}}{m_{B_{(s)}}},
\label{eq2}
\eeq
where $q = p_1 - p_2$  is the momentum of the lepton pair.
The momenta of the spectator quarks in $B_{(s)}$ and $D_{(s)}^{(*)}$ mesons are parameterized as
\beq
k_1 =\frac{m_{B_{(s)}}}{\sqrt{2}}(0,x_1,k_{1\bot}),\quad k_2=\frac{r\cdot m_{B_{(s)}}}{\sqrt{2}}(x_2\eta^+,x_2\eta^-,k_{2\bot}),
\label{eq3}
\eeq
where $x_1$ and $x_2$ are the fraction of the momentum carried by the light spectator quark  in the initial $B/B_s$ meson and  the final state
meson $D^{(*)}/D_s^{(*)}$, respectively.

For the $B/B_s$ meson wave function, we use the same one as being used in Refs.\cite{Li,bs}.
\beq
\Phi_{B_{(s)}}(x,b)&=&\frac{\rm i}{\sqrt{2N_{\rm c}}} (\psl_{B_{(s)}} +m_{B_{(s)}}) \gamma_5 \phi_{B_{(s)}} (x,b),  \label{eq:phibs}\\
\phi_{B_{(s)}}(x,b)&=& N_{B_{(s)}} \cdot x^2(1-x)^2 \cdot \mathrm{\exp}
\left[-\frac{x^2 \cdot M_{B_{(s)}}^2}{2 \omega_{B_{(s)}}^2} -\frac{1}{2} (\omega_{B_{(s)}} \cdot b)^2\right].
\label{eq:ampbbs}
\eeq
The normalization factor $N_{B_{(s)}}$ depends on the values of the parameter $\omega_{B_{(s)}}$ and decay
constant $f_{B_{(s)}}$ through the normalization relation: $\int_0^1{\rm d}x\ \phi_{B_{(s)}}(x,b=0)=f_{B_{(s)}}/(2\sqrt{6})$.
In order to estimate the uncertainties of theoretical predictions, we set the shape parameter $\omega_B = 0.40 \pm 0.04$ GeV
and $\omega_{B_s} = 0.50 \pm 0.05$ GeV.

For the $D_{(s)}$ and $D_{(s)}^*$ meson, we use the same wave functions as being used in Ref.~\cite{ds}
\beq
\Phi_{D_{(s)}}(p,x)&=&\frac{\rm i}{\sqrt{6}}\gamma_5 (\psl_{D_{(s)}}+ m_{D_{(s)}} )\phi_{D_{(s)}}(x),   \label{eq:phid01} \\
\Phi_{D^*_{(s)}}(p,x) &=& \frac{- \rm i}{\sqrt{6}} \left
 [  \epsl_{\rm L}(\psl_{D^*_{(s)}} +m_{D^*_{(s)}})\phi^{\rm L}_{D^*_{(s)}}(x)
 + \epsl_{\rm T}(\psl_{D^*_{(s)}} + m_{D^*_{(s)}})\phi^{\rm T}_{D^*_{(s)}}(x)\right].
\label{eq:phids01}
 \eeq
with the distribution amplitudes
\beq
\phi_{ D^{(*)}_{(s)} }(x)=\frac{ f_{D^{(*)}_{(s)} }}{2\sqrt{6}} \cdot 6x(1-x)
\left[ 1+C_{D^{(*)}_{(s)} }(1-2x)\right] \cdot \exp\left[-\frac{\omega^2 b^2 }{2}\right].
\label{eq:phid02}
\eeq
where we set the parameters $C_D=C_{D^*} =C_{D_s}=C_{D_s^*}=0.5$ and $\omega =0.1$.
From the heavy quark limit, we here assume that
\beq
f^{L}_{D^*}&=&f^{T}_{D^*}=f_{D^*},   \quad f^{L}_{D_S^*}=f^{T}_{D_S^*}=f_{D_s^*},  \label{eq:flt1}\\
\phi^{L}_{D^*}&=&\phi^{T}_{D^*}=\phi_{D^*}, \quad   \phi^{L}_{D_s^*}= \phi^{T}_{D_s^*}=\phi_{D_s^*} .
\label{eq:philt1}
\eeq

\subsection{Form Factors in PQCD approach}

The form factors of the $B_{(s)} \to D_{(s)}$ transition are defined in the same form as in Ref.~\cite{ff}
\beq
\langle D_{(s)}(p_2)|\bar{c}(0)\gamma_{\mu}b(0)|B_{(s)}(p_1)\rangle &=&
\left[(p_1+p_2)_{\mu}-\frac{m_{B_{(s)}}^2-m_{D_{(s)}}^2}{q^2}q_{\mu}\right] F_+(q^2)
\non &+& \left[ \frac{m_{B_{(s)}}^2-m_{D_{(s)}}^2}{q^2}q_{\mu} \right] F_0(q^2) \
\label{eq:fpf0}
\eeq
In order to cancel the poles at $q^2 = 0$, $F_+(0)$ should be equal to $F_0(0)$. For convenience, we also define the auxiliary
form factors $f_1(q^2)$ and $f_2(q^2)$,
\beq
\langle D_{(s)}(p_2)|\bar{c}(0)\gamma_{\mu}b(0)|B_{(s)}(p_1)\rangle=f_1(q^2)p_{1\mu}+f_2(q^2)p_{2\mu},
\label{eq:f1f2}
\eeq
where the auxiliary form factors $f_1(q^2)$ and $f_2(q^2)$ are related to $F_+(q^2)$ and $F_0(q^2)$ through the following relations,
\beq
F_+(q^2)&=&\frac12\left[f_1(q^2)+f_2(q^2)\right], \label{eq:fp01}\\
F_0(q^2)&=&\frac12 f_1(q^2)\left[1+\frac{q^2}{m_{B_{(s)}}^2-m_{D_{(s)}}^2}\right]
+\frac12 f_2(q^2)\left[1-\frac{q^2}{m_{B_{(s)}}^2-m_{D_{(s)}}^2}\right].
\label{eq:f001}
\eeq

As for the vector meson $D^*_{(s)}$ at the final state, the form factors involved in $B_{(s)} \to D^*_{(s)}$ transitions are
$V(q^2)$ and $A_{0,1,2}(q^2)$.  And they are defined in the following forms \cite{ff} :
\beq
\langle D^*_{(s)}(p_2)|\bar{c}(0) \gamma_{\mu} b(0)|B_{(s)}(p_1)\rangle
&=& \frac{2 i V(q^2)}{m_{B_{(s)}}+m_{D_{(s)}^*}}\epsilon_{\mu \nu \alpha
\beta} \epsilon^{* \nu}p_1^\alpha p_2^\beta,\\
\langle D^*_{(s)}(p_2)|\bar{c}(0) \gamma_{\mu}\gamma_5 b(0)|B_{(s)}(p_1)\rangle
&=& 2 m_{D_{(s)}^*}A_0(q^2)\frac{\epsilon^*\cdot q}{q^2}q_\mu \non
&& \hspace{-2cm}+ (m_{B_{(s)}} + m_{D_{(s)}^*})A_1(q^2) \left (\epsilon^*_\mu - \frac{\epsilon^*\cdot q}{q^2}q_\mu \right )\non
& & \hspace{-2cm} - A_2(q^2)\frac{\epsilon^*\cdot q}{m_{B_{(s)}} + m_{D_{(s)}^*}} \left [(p_1+p_2)_\mu -
\frac{m_{B_{(s)}}^2-m_{D_{(s)}^*}^2}{q^2}q_\mu \right ].
\label{eqff2} \eeq

We calculated the relevant form factors mentioned above in PQCD approach, and the analytical expressions are like follows:
\beq
 f_1(q^2)&=&8\pi m^2_{B_{(s)}} C_F\int dx_1 dx_2\int b_1 db_1 b_2  db_2 \phi_{B_{(s)}}(x_1,b_1) \phi_{D_{(s)}}(x_2,b_2)\non
&\times & \Bigl\{ \left[ 2 r\left(1-rx_2\right)  \right]\cdot H_1(t_1) \non
&+& \left[ 2 r(2 r_c-r)
 + x_1 r \left (-2+2 \eta+\sqrt{\eta^2-1}-\frac{2\eta}{\sqrt{\eta^2-1}}+\frac{\eta^2}{\sqrt{\eta^2-1}} \right ) \right] \cdot H_2(t_2) \Bigr \},\ \
\label{eqf1} \eeq
\beq
f_2(q^2)&=&8\pi m^2_{B_{(s)}} C_F\int dx_1 dx_2\int b_1 db_1 b_2 db_2
\phi_{B_{(s)}}(x_1,b_1)\phi_{D_{(s)}}(x_2,b_2)\non
&\times&  \Bigl\{ \left[ 2-4 x_2 r(1-\eta) \right] \cdot H_1(t_1)
+ \left[ 4r-2r_c-x_1+\frac{x_1}{\sqrt{\eta^2-1}}(2-\eta) \right] \cdot H_2(t_2) \Bigr \}, \ \ \
\label{eqf2} \eeq
\beq
V(q^2)&=&8\pi m^2_{B_{(s)}} C_F\int dx_1 dx_2\int b_1 db_1 b_2 db_2
\phi_{B_{(s)}}(x_1,b_1)\phi^T_{D^*_{(s)}}(x_2,b_2) \cdot (1+r)\non
&\times & \Bigl \{\left[1-rx_2\right] \cdot H_1(t_1) + \left[r+\frac{x_1}{2\sqrt{\eta^2-1}}\right] \cdot H_2(t_2) \Bigr \},
\label{eqV} \eeq
\beq
A_0(q^2) &=& 8\pi m^2_{B_{(s)}} C_F\int dx_1 dx_2\int b_1 db_1 b_2 db_2 \phi_{B_{(s)}}(x_1,b_1)\phi^L_{D^*_{(s)}}(x_2,b_2)\non
&\times & \Bigl \{ \left[ 1+r -rx_2(2+r-2\eta)\right]\cdot H_1(t_1) \non
&+& \left [r^2+r_c+\frac{x_1}{2}+\frac{\eta x_1}{2\sqrt{\eta^2-1}}
+\frac{rx_1}{2\sqrt{\eta^2-1}}(1-2\eta(\eta+\sqrt{\eta^2-1}))\right ]\cdot H_2(t_2) \Bigr \},\ \ \
\label{eqA0} \eeq
\beq
A_1(q^2)&=&8\pi m^2_{B_{(s)}} C_F\int dx_1 dx_2\int b_1 db_1 b_2 db_2
\phi_{B_{(s)}}(x_1,b_1)\phi^T_{D^*_{(s)}}(x_2,b_2)\cdot \frac{r}{1+r}\non
&\times & \Bigl \{2 [ 1+\eta-2 r x_2+r\eta x_2 ]\cdot H_1(t_1) + \left[2r_c+2 \eta r-x_1\right]\cdot H_2(t_2) \Bigr \},
\label{eqA1}
\eeq
\beq
A_2(q^2) &=& \frac{(1+r)^2(\eta-r)}{2r(\eta^2-1) }\cdot A_1(q^2) \non
&-&  8\pi m^2_{B_{(s)}} C_F\int dx_1 dx_2\int b_1 db_1 b_2 db_2\phi_{B_{(s)}} (x_1,b_1) \cdot \phi^L_{D^*_{(s)}}(x_2,b_2) \cdot \frac{1+r}{\eta^2-1} \non
&\times & \Bigl \{  \left[(1+\eta)(1-r) -rx_2(1-2r+\eta(2+r-2\eta))\right] \cdot H_1(t_1) \non
&+& \left[ r+r_c(\eta-r)-\eta r^2+r x_1\eta^2-\frac{x_1}{2}(\eta+r)+x_1 (\eta r
-\frac{1}{2})\sqrt{\eta^2-1}\right] \cdot H_2(t_2) \Bigr \},
\label{eqA2}
\eeq
where the color factor $C_F=4/3$, the mass ratio $r_c=m_c/m_{B_{(s)}}$, and  the functions $H_i(t_i)$ are in the following form
\beq
H_i(t_i)=h_i(x_1,x_2,b_1,b_2) \cdot \alpha_s(t_i) \exp\left [-S_{ab}(t_i) \right ], \quad {\rm  for} \quad   i=(1,2) .  \label{eqhiti}
\eeq
The explicit expressions of the hard functions $h_i(x_1,x_2,b_1,b_2)$, the hard scales $t_i$ and the Sudakov factors $ [-S_{ab}(t_i)] $ will be given
in the Appendix.

\subsection{Lattice inputs at  $q^2_{max}$}

The lattice QCD has its own advantages to calculate the relevant form factors at large $q^2$ region.
We generally believe that the lattice QCD predictions for those relevant form factors  close or at the $q^2_{max}$ are reliable.
In this work, we make use of this reliability and take the lattice QCD predictions for all relevant form factors at the endpoint $q^2_{max}$
as the additional inputs in the fitting process so that the extrapolation of these form factors from the low $q^2$ region to $q^2_{max}$
could become  more reliable.

The form factors used in the lattice QCD are parameterized as follows \cite{laff1,laff2}:
\beq
\langle D_{(s)}|\bar{c} V^{\mu} b|B_{(s)}\rangle &=& \sqrt{m_{B_{(s)}} m_{D_{(s)}}}
\left[ h_+(w)(\upsilon+\upsilon')^{\mu}+h_-(w)(\upsilon-\upsilon')^{\mu} \right] , \non
\langle D^*_{(s)}|\bar{c} V^{\mu} b|B_{(s)}\rangle &=& \sqrt{m_{B_{(s)}} m_{D^*_{(s)}}} i\varepsilon^{\mu\nu}_{~~~\rho\sigma} \epsilon^*_\nu
\upsilon^\rho \upsilon'^\sigma h_V(w) , \non
\langle D^*_{(s)}|\bar{c} A^{\mu} b|B_{(s)}\rangle &=& \sqrt{m_{B_{(s)}} m_{D^*_{(s)}}} \epsilon^*_\nu
\left[ g^{\mu\nu}(1+w)h_{A_1}(w)-\upsilon^\nu(\upsilon^\mu h_{A_2}(w)+\upsilon'^\mu h_{A_3}(w))\right],
\label{eqlattice} \eeq
where $\upsilon=p_{B_{(s)}}/m_{B_{(s)}}, \upsilon'=p_{D^{(*)}_{(s)}}/m_{D^{(*)}_{(s)}}$, the velocity transfer $w = \upsilon \cdot \upsilon'$,
and $\epsilon$ is the polarization vector of the $D^*_{(s)}$ meson. Through a simple transformation, we can relate them to the form factors used in our work:
\beq
F_+(q^2)&=& \frac{1}{2\sqrt{r}} \left [ (1+r) h_+(w) - (1-r)h_-(w) \right] ,\non
F_0(q^2) &=& \sqrt{r} \left [ \frac{1+w}{1+r}h_+(w) -\frac{w-1}{1-r} h_-(w) \right] ,
\label{eqtran1} \eeq
where $w= (m^2_{B_{(s)}}+m^2_{D_{(s)}} -q^2)/(2 m_{B_{(s)}} m_{D_{(s)}})$, and
\beq
V(q^2)&=&\frac{1+r}{2\sqrt{r}}h_V(w),\non
A_0(q^2)&=&\frac{1}{2\sqrt{r}}\left [ (1+w)h_{A_1}(w)-(1-w r)
h_{A_2}(w)+(r-w)h_{A_3}(w) \right ],\non
A_1(q^2)&=&\frac{\sqrt{r}}{1+r}(1+w)h_{A_1}(w),\non
A_2(q^2)&=&\frac{1+r}{2\sqrt{r}}(r h_{A_2}(w)+h_{A_3}(w)),
\label{eqtran2} \eeq
where $w= (m^2_{B_{(s)}}+m^2_{D^*_{(s)}} -q^2)/(2 m_{B_{(s)}} m_{D^*_{(s)}})$.

At the end point $q^2=q^2_{max}$, we have
\beq
h_V(1)&=& h_{A_1}(1)=h_{A_3}(1), \quad h_{A_2}(1)=0, \quad h_{A_1}(1)={\cal F}(1), \non
w&=&1, \quad \left [ h_+(1) - \frac{(1-r)}{(1+r)}h_-(1) \right]= {\cal G}(1).
\eeq
Therefore, those relevant form factors at the endpoint  $q^2_{max}$ will  become:
\beq
F_+(q^2_{max}) &=& \frac{1+r}{2\sqrt{r}} {\cal G}(1) , \non
V(q^2_{max})&=&A_0(q^2_{max})=A_2(q^2_{max})=\frac{1}{A_1(q^2_{max})} = \frac{1+r}{2\sqrt{r}} {\cal F}(1),
\label{eqtran3}
\eeq
By using the formulaes above and including the lattice inputs \cite{lattice5,lain}
\beq
{\cal G}^{B \to D}(1)=1.033\pm 0.095,  \quad {\cal F}^{B \to D^*}(1)=0.895\pm0.010\pm0.024, \non
{\cal G}^{B_s \to D_s}(1)=1.052 \pm 0.046 ,  \quad  {\cal F}^{B_s \to D_s^*}(1)=0.883\pm0.012\pm0.028,
\label{eqinput} \eeq
with the relation between $F_0$ and $F_+$ near  the endpoint $q^2_{max}$  as evaluated  in Ref.~\cite{lain}:
\beq
F_0^{B \to D}/F_+^{B \to D}=0.73\pm0.04, \quad F_0^{B_s \to D_s}/F_+^{B_s \to D_s}=0.77\pm0.02,
\label{eqff3} \eeq
We find  the following values of the relevant form factors at the endpoint $q^2_{max}$:
\beq
F^{B \to D}_0(q^2_{max})&=& 0.86 \pm 0.08,~~ F^{B_s \to D_s}_0(q^2_{max})= 0.91 \pm 0.05, \non
F^{B \to D}_+(q^2_{max})&=& 1.17 \pm 0.10,~~ F^{B_s \to D_s}_+(q^2_{max})= 1.19 \pm 0.05, \non
V^{B \to D^*}(q^2_{max})&=& 1.01 \pm 0.05,~~ V^{B_s \to D_s^*}(q^2_{max})= 0.98 \pm 0.05, \non
A^{B \to D^*}_0(q^2_{max})&=& 1.01 \pm 0.05,~~ A^{B_s \to D_s^*}_0(q^2_{max})= 0.98 \pm 0.05, \non
A^{B \to D^*}_1(q^2_{max})&=& 0.80 \pm 0.04,~~ A^{B_s \to D_s^*}_1(q^2_{max})= 0.79 \pm 0.04, \non
A^{B \to D^*}_2(q^2_{max})&=& 1.01 \pm 0.05,~~ A^{B_s \to D_s^*}_2(q^2_{max})= 0.98 \pm 0.05.
\label{eq:ffvalue1}
\eeq
The uncertainty of the form factors comes from the errors of ${\cal G}(1)$ as given in Eq.~(\ref{eqinput}) is around
$5\sim10\%$, while the errors of ${\cal F}(1)$ is around 2 percent only.
We here set conservatively the common error of $5\%$ for the form factors $V,A_{0,1,2}$ in Eq.~(\ref{eq:ffvalue1}) by taking into account
approximately the small variations of the central values of ${\cal G}(1)$ and ${\cal F}(1)$ in recent years \cite{lattice5,laff1,laff2,lain} .

\subsection{Extrapolations and Differential decay rates}

We know that the form factors calculated by the PQCD approach are only reliable at the low $q^2$ region.
In order to cover the whole momentum region $m_l^2\leq q^2 \leq  q^2_{max}$ we have to do  the extrapolations, and
we also use the values of the relevant form factors at the endpoint $q^2_{max}$ evaluated based on the lattice QCD
for the purpose of improving the extrapolation.

In the previous works \cite{pqcd1,pqcd2}, we used the pole model parametrization \cite{polemodel} to do the fitting.
In this work,  we  will use the  Bourrely-Caprini-Lellouch (BCL) parametrization \cite{bcl09} instead of  the pole model one,
in order to match the lattice inputs and  improve the reliability of the extrapolation of the form factors from the low $q^2$ region to $q^2_{max}$.
We use our PQCD predictions for all relevant form factors $f_i(q^2)$
at the sixteen points of $0\leq q^2 \leq m^2_\tau$ as inputs and the lattice QCD results as additional inputs at the endpoints $q^2_{max}$,
then make the extrapolation from  the low $q^2$ region  to the endpoint $q_{\rm max}^2$ by using  the
BCL parametrization \cite{bcl09}.
Analogous to Ref.~\cite{jhep1905-094}, we here also consider only the first two terms  of the series  in the parameter $z$:
\begin{eqnarray}
f_i(t) = \frac{1}{1-t/m^2_R} \sum_{k=0}^{1} \alpha^i_k\; z^k(t,t_0)
=\frac{1}{1-t/m^2_R} \left ( \alpha^i_0 + \alpha^i_1\;    \frac{ \sqrt{t_+ - t} - \sqrt{t_+ - t_0}}{\sqrt{t_+ - t} + \sqrt{t_+ - t_0}} \right ) ,
\label{eq:extra1}
\end{eqnarray}
with
\beq
0 \leq t_{0} &=&  t_+  \left (1 - \sqrt{1 - \frac{t_-}{t_+}} \right ) \leq t_-,   \label{eq:t0}
\eeq
where $t=q^2$,  $t_{\pm} = (m_{B_{(s)}} \pm m_{D^{(*)}_{(s)}} )^2$ , and  $m_R$ are the masses of the low-lying resonance.
The optimized value of $t_0$  and the values of $m_R$ are chosen to be  the same ones as in Ref.~\cite{jhep1905-094}.
Since the choice of $m_R$ depends on the kinds of the charged current  involved for the considered semileptonic decays, i.e.
the $b \to c l\bar{\nu}_l$ or  the $b \to u l \bar{\nu}_l$ transition, we use the same set of $m_R$ as that in Ref.~\cite{jhep1905-094}
where the $B_c \to (\eta_c,\jpsi) l \bar{\nu}_l$ decays  had the same quark level $b \to c l\bar{\nu}_l$  transitions  as  this paper.

With the extrapolations above, now we have the access to the full momentum dependence of those relevant form factors, and the branching ratios of
the semileptonic decays $B_{(s)} \to D^{(*)}_{(s)} l \nu$ can be calculated.
The quark level transition of these semileptonic decays is the $b\to c l \nu$ transition with the effective Hamiltonian \cite{rmp68-1125}
\beq
{\cal H}_{eff}(b\to c l \nu)=\frac{G_F}{\sqrt{2}}V_{cb}\;
\bar{c} \gamma_{\mu}(1-\gamma_5)b \cdot \bar l\gamma^{\mu}(1-\gamma_5)\nu_l.
\label{eq-hamiltonian} \eeq
where $G_F=1.16637\times10^{-5} GeV^{-2}$ is the Fermi-coupling constant. And the differential decay rates of the decay mode
$B_{(s)} \to D_{(s)} l \nu$ can be expressed as \cite{prd79-014013}:
\beq
\frac{d\Gamma(B_{(s)} \to D_{(s)} l \nu)}{dq^2}&=&\frac{G_F^2|V_{cb}|^2}{192 \pi^3  m_{B_{(s)}}^3}
\left ( 1-\frac{m_l^2}{q^2} \right)^2\frac{
\lambda^{1/2}(q^2)}{2q^2}\cdot \Bigl \{  3 m_l^2 \left ( m_{B_{(s)}}^2- m^2_{D_{(s)}}  \right )^2 |F_0(q^2)|^2 \non &&
+ \left ( m_l^2+2q^2 \right )\lambda(q^2) |F_+(q^2)|^2 \Bigr \},
\label{eq:dg1} \eeq
where $m_l$ is the mass of the leptons $e$, $\mu$ or $\tau$,  and
$\lambda(q^2) = (m_{B_{(s)}}^2+m_{D_{(s)}}^2-q^2)^2 - 4 m_{B_{(s)}}^2 m_{D_{(s)}}^2$ is the phase space factor.

For the decay mode $B_{(s)} \to D^*_{(s)} l \nu$ , the corresponding differential decay widths
can be written as ~\cite{polemodel}:
\beq
\frac{d\Gamma_L(B_{(s)} \to D^*_{(s)} l \nu)}{dq^2}&=& \frac{G_F^2|V_{cb}|^2}{192 \pi^3  m_{B_{(s)}}^3}
\left ( 1-\frac{m_l^2}{q^2}\right )^2
\frac{\lambda^{1/2}(q^2)}{2q^2}\cdot \Bigg\{3m^2_l\lambda(q^2)A^2_0(q^2)\non
&& \hspace{-3.5cm}
+\frac{m^2_l+2q^2}{4m^2} \left [ \left ( m^2_{B_{(s)}}-  m^2_{D_{(s)}^*}-q^2 \right ) \left ( m_{B_{(s)}}+  m_{D_{(s)}^*}  \right )A_1(q^2)
 -\frac{\lambda(q^2)  A_2(q^2)   }{ m_{B_{(s)}} +  m_{D_{(s)}^* } } \right ]^2 \Bigg\},\ \ \
\label{eq:dfds1}
\eeq
\beq
\frac{d\Gamma_T(B_{(s)} \to D^*_{(s)} l \nu)}{dq^2}&=& \frac{G_F^2|V_{cb}|^2}{192 \pi^3 m_{B_{(s)}}^3}
\left ( 1-\frac{m_l^2}{q^2}\right )^2  \lambda^{3/2}(q^2)  \left ( m^2_l+2q^2 \right ) \non
& \times & \left [ \frac{V^2(q^2)}{ \left ( m_{B_{(s)}}+ m_{D_{(s)}^*}  \right )^2 } +  \frac{ \left  (m_{B_{(s)}}+ m_{D_{(s)}^*}  \right )^2 A^2_1(q^2)}{ \lambda(q^2)} \right ],
\label{eq:dfds2}
\eeq
with the phase space factor $\lambda(q^2) = (m_{B_{(s)}}^2+m_{D^*_{(s)}}^2-q^2)^2 - 4 m_{B_{(s)}}^2 m_{D^*_{(s)}}^2$.
The total differential decay widths are defined as:
\beq
\frac{d\Gamma}{dq^2}=\frac{d\Gamma_L}{dq^2} +\frac{d\Gamma_T}{dq^2} \; .
\label{eq:dfdst}
\eeq

\subsection{Additional physical  observables: $P_\tau$, $F_L(D_{(s)}^*)$ and  $A_{FB}(\tau)$}

Besides the decay rates and the ratios $R(X)$, there are other three additional physical observables for the considered
$B/B_s$ semileptonic decays:  the longitudinal polarization of the tau lepton $P_\tau$,  the fraction of $D_{(s)}^*$
longitudinal polarization  $F_L(D_{(s)}^*)$ and the forward-backward asymmetry of the tau lepton $A_{FB}(\tau)$.
These additional physical observables are sensitive to some kinds of new physics beyond the SM \cite{prd072012,prd114022,prd036021}.

As for the definitions of these additional physical obsevables, we follow Refs.~\cite{ptau1,ptau2,ptau3}.
For the $\tau$ longitudinal polarization,  for example, the authors of Refs.~\cite{ptau1,ptau2} have studied them in the
${\bf Q}$ rest frame where the spacial components of the momentum transfer ${\bf Q }= {\bf  p}_B-{\bf  p }_M$ vanish.
They used a coordinate system in the ${\bf Q}$ rest frame so that the direction
of the $B$ momenta is along the $z$ axis, and the $\tau$ momentum lies in the $x-z$ plane.
Here, we use the same definition:
\beq
P_\tau(D_{(s)}^{(*)})=\frac{\Gamma_+(D_{(s)}^{(*)}) - \Gamma_- (D_{(s)}^{(*)})}{\Gamma_+ (D_{(s)}^{(*)})
+ \Gamma_- (D_{(s)}^{(*)})},  \label{eq:ptau}
\eeq
where $\Gamma_\pm(D_{(s)}^{(*)})$ denotes the decay rate of the decay $B_{(s)} \to D_{(s)}^{(*)}\tau \bar{\nu}_\tau$ with the $\tau$ lepton helicity
of $\pm 1/2$.  The explicit expressions of $\Gamma_{\pm}$ in the SM can be found for example in the Appendix of Ref.~\cite{ptau3}.
For $B_{(s)} \to D_{(s)} \tau \bar{\nu}_{\tau}$ decays, we find
\beq
{d\Gamma_{+} \over dq^2} &=& {G_F^2 |V_{cb}|^2 \over 192\pi^3 m_{B_{(s)}}^3} q^2 \sqrt{\lambda(q^2)}
\left( 1 - {m_\tau^2 \over q^2} \right)^2 {m_\tau^2 \over 2q^2} \left( H_{V,0}^{s\,2} + 3 H_{V,t}^{s\,2} \right) ,  \label{eq:dga0a}\\
{d\Gamma_{-} \over dq^2} &=& {G_F^2 |V_{cb}|^2 \over 192\pi^3 m_{B_{(s)}}^3} q^2 \sqrt{\lambda(q^2)}
\left( 1 - {m_\tau^2 \over q^2} \right)^2 \left( H_{V,0}^{s\,2} \right) . \label{eq:dga0b}
\eeq
For $B_{(s)} \to D^*_{(s)} \tau \bar{\nu}_{\tau}$ decays, we have
\beq
{d\Gamma_{+} \over dq^2} &=& {G_F^2 |V_{cb}|^2 \over 192\pi^3 m_{B_{(s)}}^3} q^2 \sqrt{\lambda(q^2)}
\left( 1 - {m_\tau^2 \over q^2} \right)^2 {m_\tau^2 \over 2q^2} \left( H_{V,+}^2 + H_{V,-}^2 + H_{V,0}^2 +3 H_{V,t}^2 \right) ,  \label{eq:dga0c}\\
{d\Gamma_{-} \over dq^2} &=& {G_F^2 |V_{cb}|^2 \over 192\pi^3 m_{B_{(s)}}^3} q^2 \sqrt{\lambda(q^2)}
\left( 1 - {m_\tau^2 \over q^2} \right)^2 \left( H_{V,+}^2 + H_{V,-}^2 + H_{V,0}^2 \right) .  \label{eq:dga0d}
\eeq
where the functions $(H_{V,0}^s,H_{V,t}^s,H_{V,\pm},H_{V,0},H_{V,t})$ can be written as the combinations of the six form factors  as defined
in Eqs.~(\ref{eq:fp01},\ref{eq:f001},\ref{eqf1}-\ref{eqA2}):
\beq
H_{V,0}^s(q^2) & =& \sqrt{\lambda(q^2) \over q^2} F_+(q^2) ,  \label{eq:hvs1}\\
H_{V,t}^s(q^2) & =& {m^2_{B_{(s)}}-m^2_{D_{(s)}} \over \sqrt{q^2}} F_0(q^2) ,   \label{eq:hvs2}  \\
H_{V,\pm}(q^2) & =& (m_{B_{(s)}}+m_{D^*_{(s)}}) A_1(q^2) \mp { \sqrt{\lambda(q^2)} \over m_{B_{(s)}}+m_{D^*_{(s)}} } V(q^2) ,   \label{eq:hvv}\\
H_{V,0}(q^2)   & =& {m_{B_{(s)}}+m_{D^*_{(s)}} \over 2m_{D^*_{(s)}}\sqrt{q^2} } \left[ -(m^2_{B_{(s)}}-m^2_{D^*_{(s)}}-q^2) A_1(q^2)
+ { \lambda(q^2) A_2(q^2) \over (m_{B_{(s)}  }+m_{D^*_{(s)}})^2 }  \right] ,   \label{eq:hvv2}  \\
H_{V,t}(q^2)   & = &-\sqrt{ \lambda(q^2) \over q^2 } A_0(q^2) .
\label{eq:hvv3}
\eeq
The phase space factor $\lambda$ is the same ones as defined in the Eqs.~(\ref{eq:dg1}-\ref{eq:dfds2}).

For the fraction of $D_{(s)}^*$ longitudinal polarization $F_L(D_{(s)}^*)$, it  is defined through the  secondary decay chain $D_{(s)}^* \to D_{(s)}\pi$
of the considered semileptonic $B/B_s$ decays.   And the $F_L(D_{(s)}^*)$ can be expressed as the same form in Ref.~\cite{ptau3}:
\beq
F_L(D_{(s)}^*)=\frac{\Gamma^0}{\Gamma^0+\Gamma^{+1} + \Gamma^{-1}},
\label{eq:pD}
\eeq
and the corresponding differential decay rates are of the following form:
\beq
{d\Gamma^{\pm1} \over dq^2} &=& {G_F^2 |V_{cb}|^2 \over 192\pi^3 m_{B_{(s)}}^3} q^2 \sqrt{\lambda(q^2)}
\left( 1 - {m_\tau^2 \over q^2} \right)^2 \left ( 1+{m_\tau^2 \over 2q^2} \right ) ( H_{V,\pm}^2 ) ,  \label{eq:dgc1}\\
{d\Gamma^{0} \over dq^2} &=&  {G_F^2 |V_{cb}|^2 \over 192\pi^3 m_{B_{(s)}}^3} q^2 \sqrt{\lambda(q^2)}
\left( 1 - {m_\tau^2 \over q^2} \right)^2 \left[ (1+{m_\tau^2 \over 2q^2}) H_{V,0}^2+ {3 \over 2}{m_\tau^2 \over q^2} H_{V,t}^2 \right] .
\label{eq:dgc2}
\eeq

As for the $\tau$ lepton forward-backward asymmetry $A_{FB}(\tau)$, it's a little complicated since it is defined in the
$\tau \bar{\nu}$ rest frame. The explicit expression is of the following form \cite{ptau3}:
\beq
A_{FB} = \frac{\int^1_0 {d\Gamma \over dcos\theta} dcos\theta - \int^0_{-1} {d\Gamma \over dcos\theta} dcos\theta}
{\int^1_{-1} {d\Gamma \over dcos\theta} dcos\theta} = \frac{\int b_\theta(q^2) dq^2}{\Gamma_{B_{(s)}}}
\eeq
where the angle $\theta$ is the angle between the 3-momenta of the lepton $\tau$ and the initial $B$ or $B_s$
in the  $\tau \bar{\nu}$ rest frame. The function $ b_\theta(q^2)$ is the angular coefficient which can be written as \cite{ptau3}:
\beq
b_\theta^{(D)}(q^2) &=& {G_F^2 |V_{cb}|^2 \over 128\pi^3 m_{B_{(s)}}^3} q^2 \sqrt{\lambda(q^2)}
\left( 1 - {m_\tau^2 \over q^2} \right)^2 {m_\tau^2 \over q^2} ( H^s_{V,0}H^s_{V,t} ) , \label{eq:btheta1}\\
b_\theta^{(D^*)}(q^2) &=& {G_F^2 |V_{cb}|^2 \over 128\pi^3 m_{B_{(s)}}^3} q^2 \sqrt{\lambda(q^2)}
\left( 1 - {m_\tau^2 \over q^2} \right)^2 \left[ {1 \over 2}(H_{V,+}^2-H_{V,-}^2)+ {m_\tau^2 \over q^2} ( H_{V,0}H_{V,t} ) \right] .
\label{eq:btheta2}
\eeq
With the above definitions and formulae, we are now ready to give our numerical results and phenomenological analysis.

\section{Numerical Results and Discussions} \label{sec:3}

In the numerical calculations, we use the following input parameters
(here masses and decay constants in units of GeV)\cite{hfag2017,hflav2019,pdg2018,ETM}:
\beq
m_{\rm B}&=&5.28,\quad m_{\rm B_s}=5.367,\quad m_{\rm D_0}=1.865,\quad m_{\rm D_+}=1.870,\quad m_{\rm D^*_0}=2.007, \non
m_{\rm D^*_+}&=&2.010,\quad m_{\rm D_s}=1.968,\quad m_{\rm D^*_s}=2.112,\quad m_{\rm \tau}=1.777,\quad m_{c}=1.275^{ \ +0.025}_{ \ -0.035}, \non
f_{\rm D}&=& 0.212, \quad f_{\rm D_s}=0.249, \quad f_{\rm B_+}=0.187, \quad f_{\rm B_0}=0.191,\quad f_{\rm B_s}=0.227, \non
|V_{\rm cb}|&=&(42.2 \pm 0.8)\times 10^{-3}, \quad \tau_{B_+}=1.638  {\rm ps},\quad \tau_{B_0}=1.520  {\rm ps}, \quad \tau_{B_s}=1.509 {\rm ps},\non
f_{\rm D^*}&=& (1.078 \pm 0.036) \cdot f_{\rm D},\quad f_{\rm D_s^*}= (1.087 \pm 0.020) \cdot f_{\rm D_s},\quad \Lambda^{\rm (f=4)}_{\overline{\rm MS}} = 0.287.
\label{eq:inputs} \eeq

\subsection{Form Factors}

For the considered  semileptonic $B/B_s$ meson decays,  it is obvious that the theoretical  predictions for the differential decay rates  and
other physical observables are strongly depended on the form factors $F_{\rm 0,+}(q^{\rm 2})$, $V(q^{\rm 2})$ and $A_{\rm 0,1,2}(q^{\rm 2})$.
To be specific, $F_{\rm 0,+}(q^{\rm 2})$ controls the process $B_{(s)} \to D_{(s)} l \nu_l$ while $V(q^{\rm 2})$ and $A_{\rm 0,1,2}(q^{\rm 2})$
play the key roles in the process $B_{(s)} \to D^*_{(s)} l \nu_l$. The value of these form factors at $q^2=0$ and their $q^2$ dependence in the whole
range of $0\leq  q^2 \leq q^2_{max}$ contain lots of information of the relevant physical process.

In Refs.~\cite{ds,Li,pqcd1}, the authors examined the applicability of the PQCD approach to $(B \to D^{(*)})$ transitions, and have shown
that the PQCD approach with the inclusion of the Sudakov effects is  applicable to the semileptonic decays $B \to D^{(*)} l\bar{\nu}_{\rm l}$
at the low $q^{\rm 2}$ region. Therefore, we present our PQCD predictions for  the relevant form factors of $B_{(s)} \to D^{(*)}_{(s)}$ transitions
at the point $q^2=0$ in Table.~\ref{table1}.

\begin{table}[thb]
\begin{center}
 \caption{  The theoretical predictions for the values of the form factors at $q^{2}=0$ using the PQCD approach with
 the BCL parametrization  [51,52]. }
\label{table1}
\vspace{0.2cm}
\begin{tabular}{l|| c| c| c| c| c} \hline\hline
\ \ \ & $F_{(0,+)}(0)$ & $V(0)$ & $A_0(0)$ & $A_1(0)$ & $A_2(0)$ \\ \hline
$B^+ \to D^0$ & $ 0.53\pm0.10 $ & $-$&$-$ &$-$ &$-$  \\
$B^0 \to D^- $ & $ 0.54\pm0.10 $ & $-$ &$-$&$-$&$-$  \\
$B^0_s \to D^-_s   $ & $ 0.52\pm0.10 $ & $-$&$-$&$-$& $-$   \\  \hline
$B^+ \to D^{*0}$ & $-$  & $ 0.64\pm0.11 $& $ 0.49\pm0.08 $& $ 0.51\pm0.09 $& $ 0.54\pm0.09 $ \\
$B^0 \to D^{*-} $ & $-$  & $ 0.65\pm0.11 $& $ 0.50\pm0.09 $& $ 0.52\pm0.08 $& $ 0.55\pm0.10 $ \\
$B^0_s   \to D^{*-}_s  $ &$-$   & $ 0.64\pm0.12 $& $ 0.48\pm0.09 $& $ 0.50\pm0.09 $& $ 0.53\pm0.11 $ \\
\hline \hline
\end{tabular} \end{center} \end{table}

From the numerical values in Table \ref{table1}, one can see easily that the same form factors corresponding to different $B \to D^{(*)}$ or $B_s \to D_s^{(*)}$
transitions  are very similar in magnitude at $q^2=0$, which implies that  the effect of the SU(3) flavor symmetry breaking is really small, less than $10\%$.
In order to cover whole $q^2$ region,  one has to make an extrapolation for all relevant form factors from the $q^2 = 0$ region to $q^2 = q^2_{max}$
region. In this work we will make the extrapolation by using two different methods.
\begin{enumerate}
\item[(1)]
The first method is analogous to the one used in Refs.~\cite{pqcd1,pqcd2}. we first calculate explicitly the values of the relevant form factors
at several points in the lower region $ 0 \leq q^{\rm 2} \leq m_{\rm \tau}^{\rm 2}$ by  using the expressions as given in Eqs.~(\ref{eqf1}-\ref{eqA2})
and the definitions in Eqs.~(\ref{eq:fp01},\ref{eq:f001}).
In the fitting process, we will use  the BCL parametrization formula as given in Eqs.~(\ref{eq:extra1},\ref{eq:t0}) instead of the
 pole model parametrization being used in Refs.~~\cite{pqcd1,pqcd2}.

\item[(2)]
The second one is the so-called  ``PQCD+Lattice'' method, similar with what we did in Ref.~\cite{pqcd3}.
Since the lattice QCD results for the form factors are reliable and accurate at $q^2 \cong q^2_{max}$ region,
we  take the lattice QCD predictions for all relevant form factors at the endpoint $q^2_{max}$ as the additional inputs in the fitting process.
In order to match the lattice inputs, we use the BCL parametrization  \cite{bcl09,jhep1905-094} to make the extrapolation.
\end{enumerate}

\begin{figure}[]
\begin{center}
\vspace{-0.5cm}
\centerline{\epsfxsize=8cm\epsffile{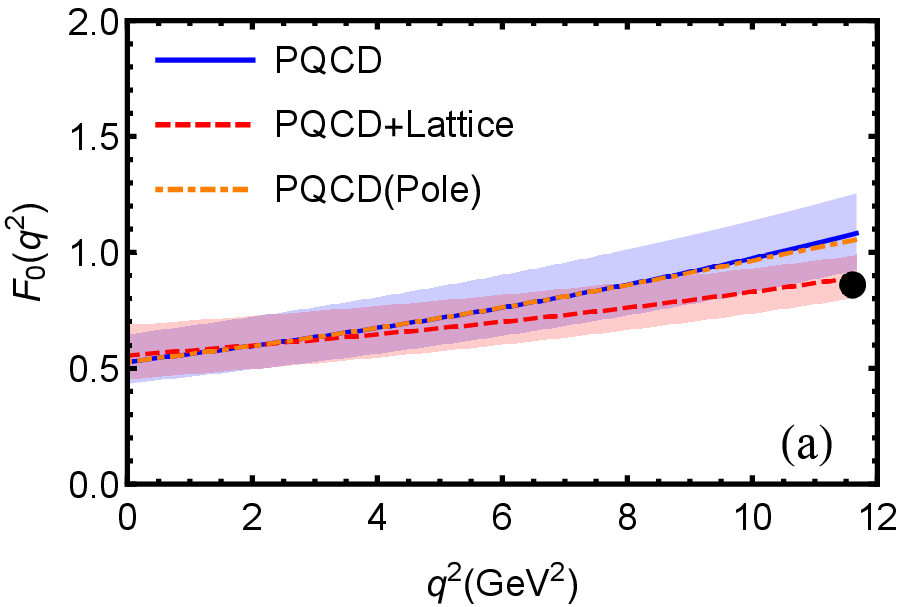}\epsfxsize=8cm\epsffile{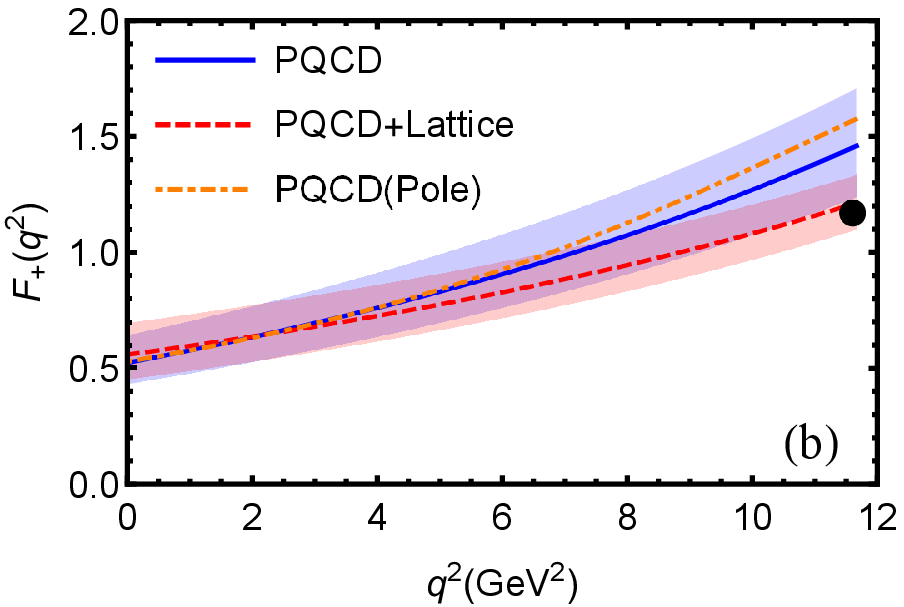} }
\vspace{0.3cm}
\centerline{\epsfxsize=8cm\epsffile{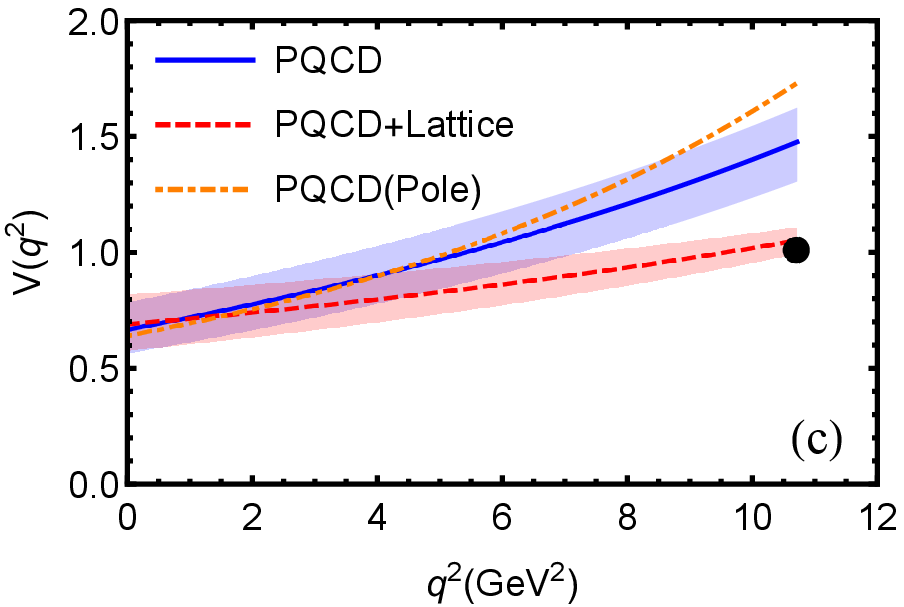}\epsfxsize=8cm\epsffile{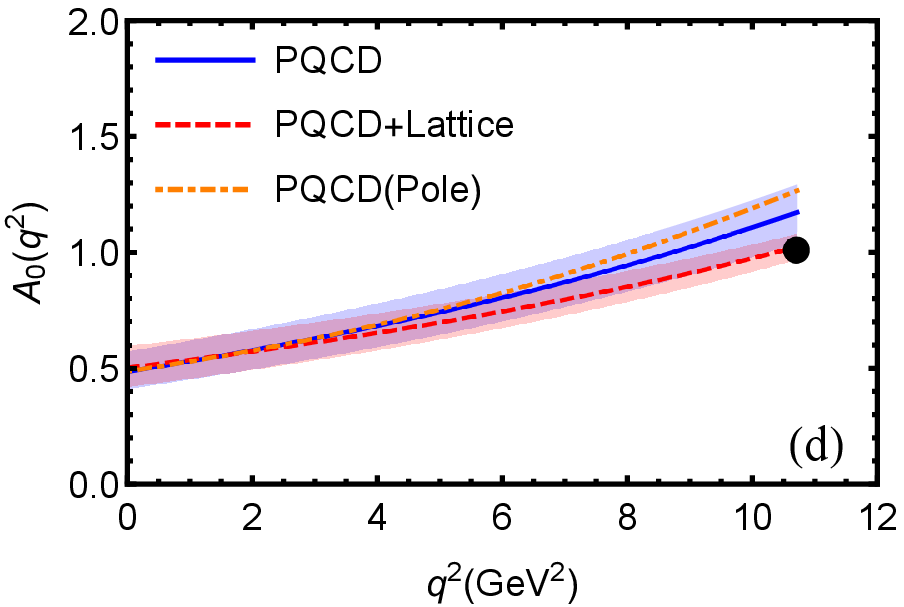} }
\vspace{0.3cm}
\centerline{\epsfxsize=8cm\epsffile{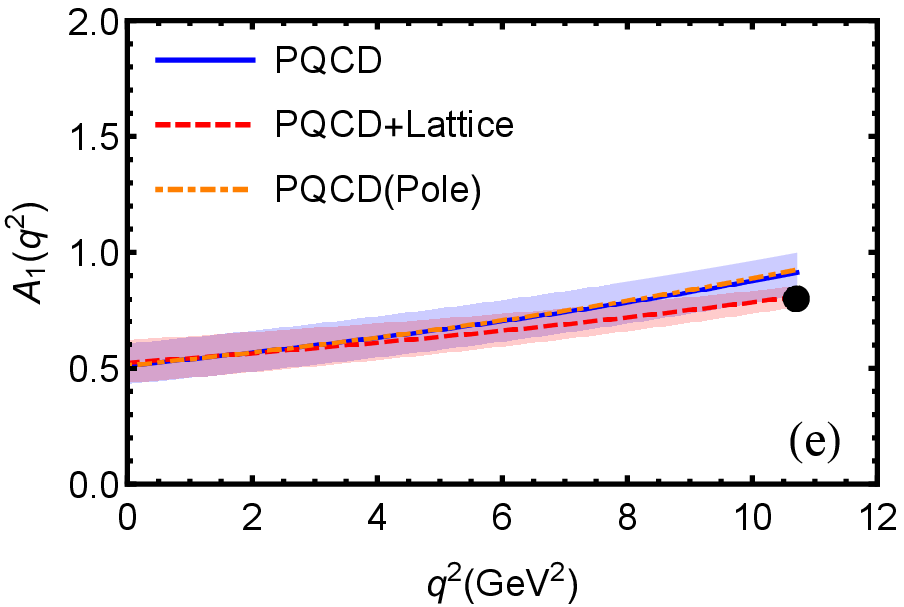}\epsfxsize=8cm\epsffile{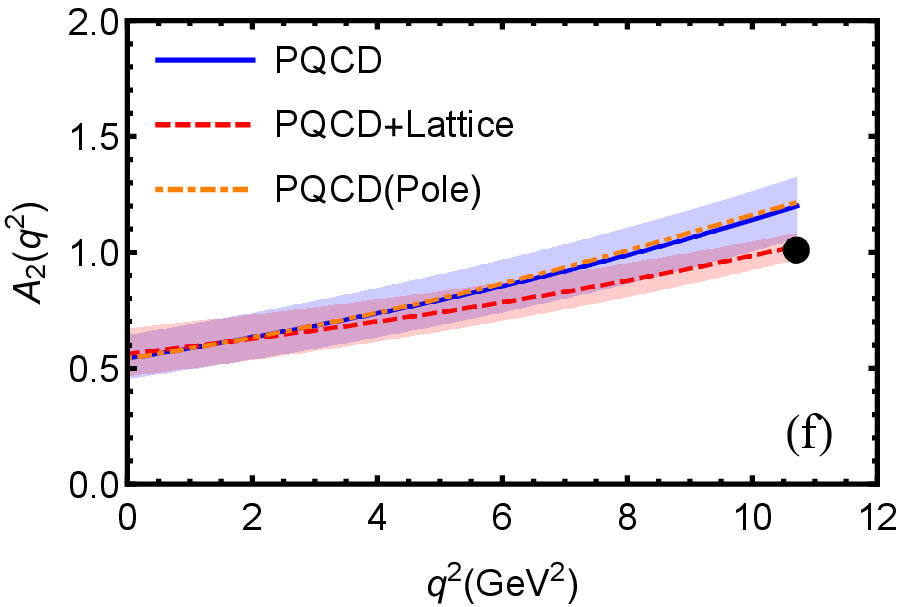} }
\caption{(Color online) The theoretical predictions for the $q^2$-dependence of the six form factors
for $B \to (D,D^{*}) $ transitions  in the PQCD approach (the blue solid curves) and the ``PQCD+Lattice'' method (the red dashed curves)
using  the BCL parametrization [51,52].
The orange dashed curves denote the PQCD predictions using the traditional pole model [9,44,60]. }
\label{fig:fig2} \end{center}
\end{figure}

\begin{figure}[]
\begin{center}
\vspace{-0.5cm}
\centerline{\epsfxsize=8cm\epsffile{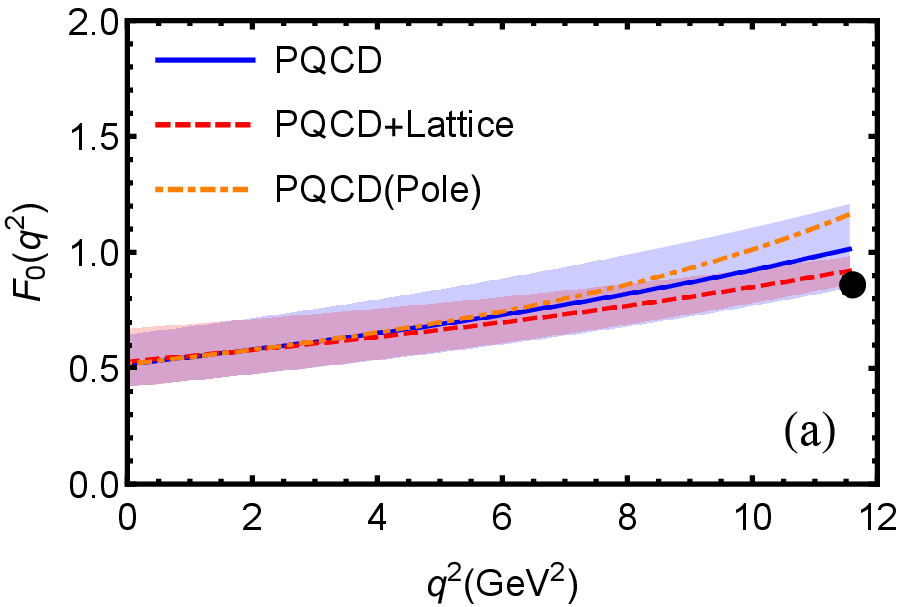}\epsfxsize=8cm\epsffile{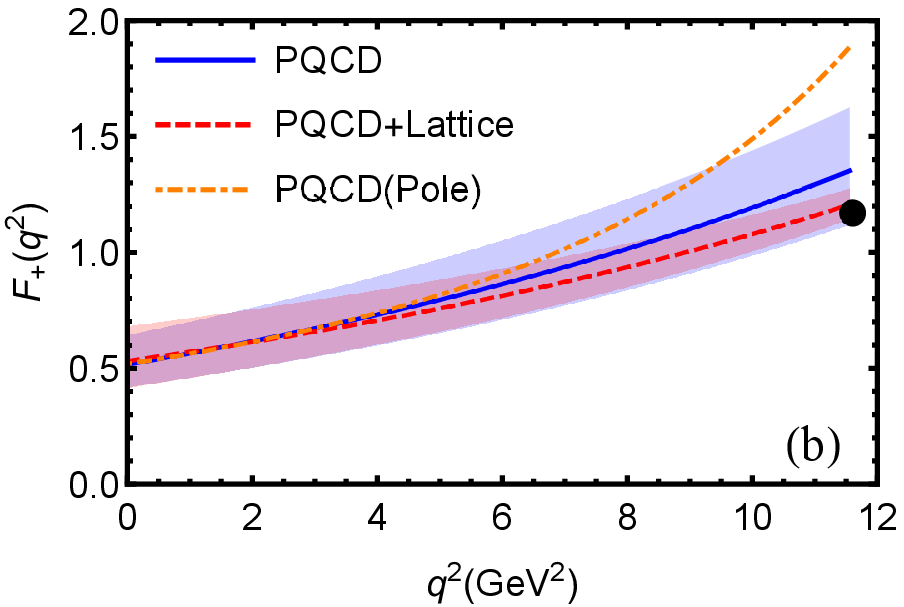} }
\vspace{0.3cm}
\centerline{\epsfxsize=8cm\epsffile{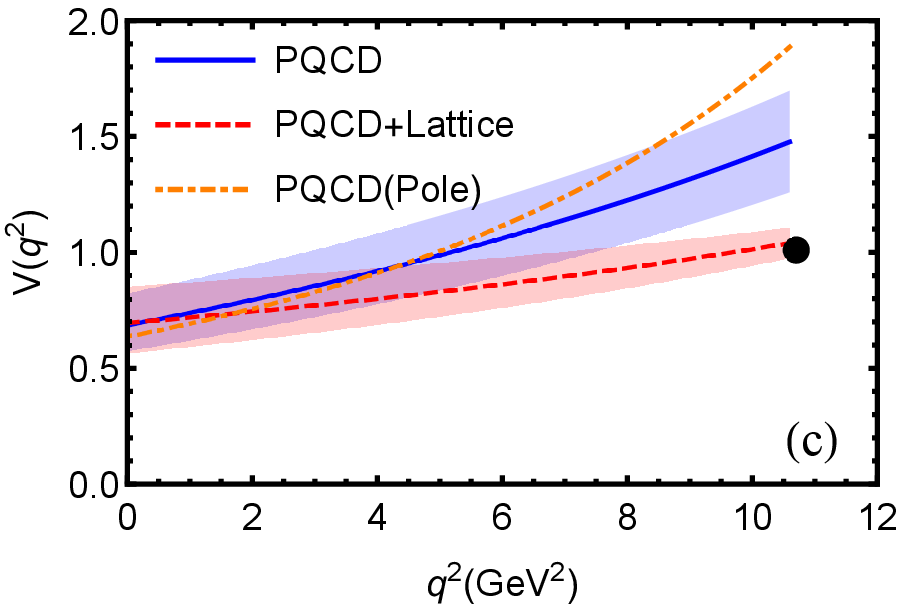}\epsfxsize=8cm\epsffile{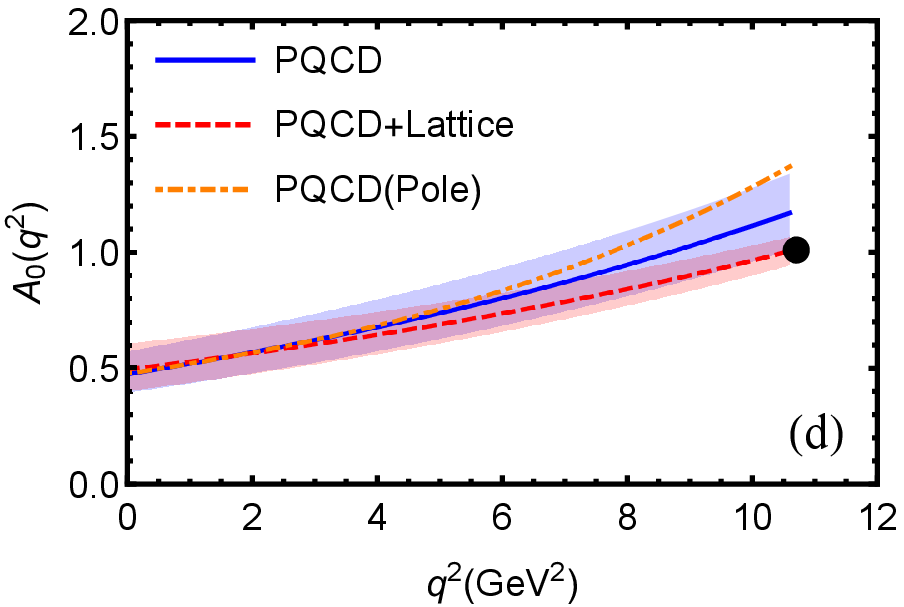} }
\vspace{0.3cm}
\centerline{\epsfxsize=8cm\epsffile{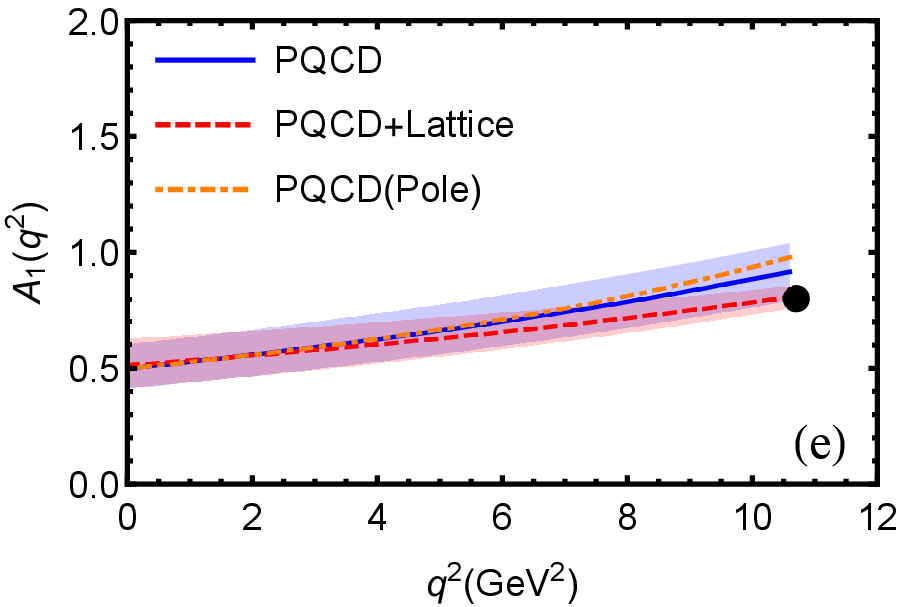}\epsfxsize=8cm\epsffile{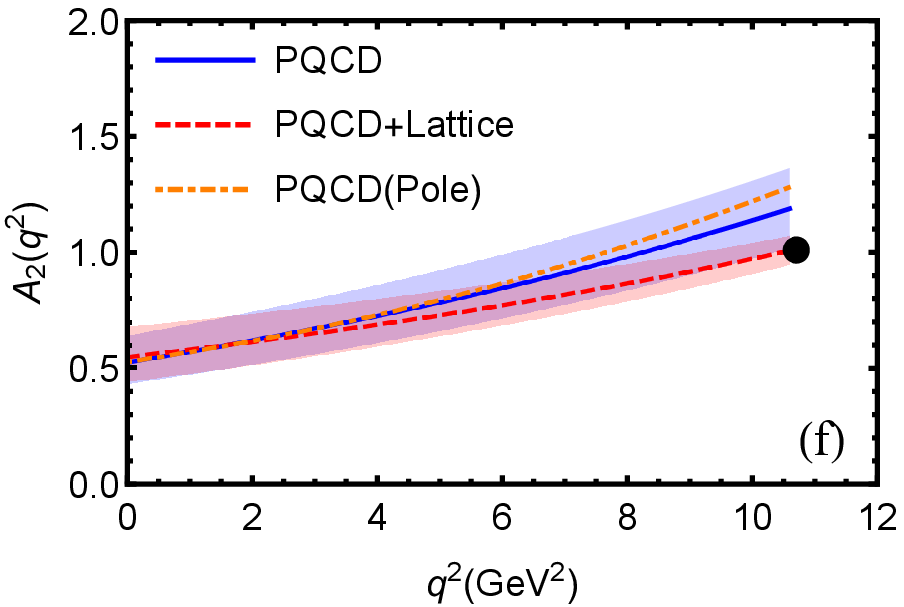} }
\caption{(Color online) The theoretical predictions for the $q^2$-dependence of the six form factors
for $B^0_s \to (D^-_s,D^{*-}_s)$ transitions in the PQCD approach (the blue solid curves)  and the ``PQCD+Lattice'' method (the red dashed curves)
using  the BCL parametrization [51,52]. The orange dashed curves denote the PQCD predictions using the
traditional pole model [9,44,60]. }
\label{fig:fig3}
\end{center}
\end{figure}

In  Figs.~\ref{fig:fig2} and \ref{fig:fig3} ,  we show the theoretical predictions for the  $q^2$ -dependence of the
six relevant form factors for $B^+ \to (D^0,D^{*0})$  and $B_s \to (D_s^-,D^{*-}_s)$  transitions,  obtained by employing
the PQCD approach and the ``PQCD + Lattice''  method.
In these two figures, the blue solid curves ( red dashed curves) show the theoretical predictions for the $q^2$-dependence of the form factors
($F_{0,+}(q^2)$, $V(q^2)$, $A_{0,1,2}(q^2)$) for the $B/B_s \to D^{(*)}/D_s^{(*)}$ transitions by using the traditional PQCD approach
(the   ``PQCD+Lattice'' approach ). The blue and red band show the theoretical uncertainties.
 As a comparison, we also show the central values of the PQCD predictions for all relevant form factors in  Figs.~\ref{fig:fig2} and \ref{fig:fig3} (the yellow dashed curves),
where the pole model parametrization ~\cite{polemodel,pqcd1,pqcd2} are used.
One can see from  the theoretical predictions as listed in Table \ref{table1} and illustrated in Figs.~\ref{fig:fig2} and \ref{fig:fig3} that
\begin{enumerate}
\item[(1)]
The form factors for the $B\to D^{(*)}$ transitions and their counterparts for $B_s\to D_s^{(*)}$ transitions
have very similar  values in  the whole $q^2$ region  due to the  SU(3) flavor  symmetry.

\item[(2)]
The differences  between the theoretical predictions of the form factors between the  traditional PQCD approach and the  "PQCD+Lattice"
method  is small in the low $q^2$ region and  remain not large even in the large $q^2$  region  for $F_{0,+}(q^2)$ and  $A_{0,1,2}(q^2)$.
For $V(q^2)$,  the difference becomes a little large in large $q^2$ region. For $B\to D^*$ transition, we found
 $V(10.71)\approx 1.53$ ($1.06$) in  the PQCD approach ( the  ``PQCD + Lattice''  method).  For  $B_s\to D_s^*$ transition,
 we found  very similar result.

\item[(3)]
The difference between the PQCD predictions of the central values of the form factors induced by using  the traditional pole model  or the BCL  model
to make the extrapolation is  very small in the region $q^2 < 8$ GeV$^2$.
The largest differences  in the region near $q^2_{max}$ are $\Delta F_+(11.55)=0.59$ (i.e,  about $31\%$ of $F_+(11.55)=1.89$ in pole model )
and $\Delta V(10.59)=0.36$ (i.e. about $19\%$ of  $V(10.59)=1.89$ in pole model ) for $B_s \to D_s^{(*)}$ transitions,
as illustrated in Fig.~\ref{fig:fig3}.

\end{enumerate}  

\subsection{Branching ratios and the ratio of Brs}

By inserting the form factors we have obtained above into the differential decay rates equations as given in
Eqs.~(\ref{eq:dg1}-\ref{eq:dfdst}),  it is straightforward to make the integration over the range of $m_l^2 \leq q^2 \leq (m_{B_{(s)}}-m_{D^{(*)}_{(s)}})^2$.
For the four semileptonic decays $B^0_s \to D^{(*)}_s \tau^+\nu_\tau$ and $B_s \to D^{(*)}_s l^+\nu_l$ with $l^+=(e^+,\mu^+)$, for example,
the theoretical predictions for their branching ratios are the following:
\beq
{\cal B}(B^0_s \to D^-_s \tau^+ \nu_\tau) &=& \left \{ \begin{array}{ll}
 0.72^{+0.32}_{-0.23}(\omega_{\rm B_s}) \pm 0.03(V_{\rm cb}) \pm 0.02 (m_{\rm c}) ,  &  {\rm PQCD}, \\
 0.63^{+0.17}_{-0.13}(\omega_{\rm B_s})\pm 0.03(V_{\rm cb}) \pm 0.02 (m_{\rm c}) ,   & {\rm PQCD+Lattice},\\
 \end{array} \right.  , \label{eq:br15}  \\
{\cal B}(B^0_s \to D^-_s l^+ \nu_l ) &=& \left \{ \begin{array}{ll}
 1.97^{+0.88}_{-0.65}(\omega_{\rm B_s}) \pm 0.08(V_{\rm cb}) \pm 0.03 (m_{\rm c}) ,   &  {\rm PQCD}, \\
 1.84^{+0.76}_{-0.50}(\omega_{\rm B_s})\pm 0.08(V_{\rm cb}) \pm 0.03 (m_{\rm c})  ,   & {\rm PQCD+Lattice}, \\
 \end{array} \right.  , \label{eq:br16}
\eeq
\beq
{\cal B}(B^0_s \to D^{*-}_s \tau^+ \nu_\tau) &=& \left \{ \begin{array}{ll}
 1.45^{+0.45}_{-0.39}(\omega_{\rm B_s}) \pm 0.06(V_{\rm cb}) \pm 0.06 (m_{\rm c}) ,  &  {\rm PQCD}, \\
 1.20^{+0.25}_{-0.22}(\omega_{\rm B_s})\pm 0.05(V_{\rm cb}) \pm 0.02 (m_{\rm c}) ,   & {\rm PQCD+Lattice},\\
 \end{array} \right.  , \label{eq:br25}  \\
{\cal B}(B^0_s \to D^{*-}_s l^+ \nu_l ) &=& \left \{ \begin{array}{ll}
 5.04^{+1.60}_{-1.41}(\omega_{\rm B_s}) \pm 0.20(V_{\rm cb}) \pm 0.16 (m_{\rm c}) ,   &  {\rm PQCD}, \\
 4.42^{+1.26}_{-0.98}(\omega_{\rm B_s})\pm 0.17(V_{\rm cb}) \pm 0.06 (m_{\rm c})  ,   & {\rm PQCD+Lattice}, \\
 \end{array} \right.  , \label{eq:br26}
\eeq
where the major theoretical errors come from the uncertainties of the input parameters
$\omega_{\rm B_s}=0.50 \pm 0.05$ GeV, $|V_{\rm cb}|=(42.2 \pm 0.8) \times 10^{-3}$
and $m_{\rm c}=1.275^{+0.025}_{-0.035}$ GeV. The possible errors from the uncertainties of the decay constants of the initial $B_s$ meson and the final
and $D_s^{(*)}$ mesons  are small and have been neglected.
In Table~\ref{tab2}, we list our PQCD and ``PQCD+Lattice" predictions for the branching ratios of the considered semileptonic decays of $B_s^0$ meson,
where the total errors  are obtained by adding the individual errors in quadrature.
As comparison,  we also show some theoretical predictions from other theoretical approaches in the framework of the SM.
Unfortunately, there  still be no experimental  results available at present.
In Table~\ref{tab3}, we show the theoretical predictions for the ratios of the branching ratios $R(D_s)$ and $R(D_s^*)$ defined in the same way as
$R(D^{(*)})$:
\beq
R(D_s)=\frac{{\cal B}(B_s^0\to D_s^-  \tau^+ \nu_\tau ) }{ {\cal B}(B_s^0\to D_s^-  l^+ \nu_l ) } =  \left \{ \begin{array}{ll}
 0.365^{+0.009}_{-0.012} ,  &  {\rm PQCD}, \\    0.341^{+0.024}_{-0.025} ,   & {\rm PQCD+Lattice},\\
 \end{array} \right.  , \label{eq:rds01}  \\
R(D^*_s)=\frac{{\cal B}(B_s^0\to D_s^{*-}  \tau^+ \nu_\tau ) }{ {\cal B}(B_s^0\to D_s^{*-}  l^+ \nu_l ) } =  \left \{ \begin{array}{ll}
 0.287^{+0.008}_{-0.011} ,  &  {\rm PQCD}, \\    0.271^{+0.015}_{-0.016} ,   & {\rm PQCD+Lattice},\\
 \end{array} \right.  , \label{eq:rds02}
\eeq
where $l$ denotes an electron or a muon.

\begin{table}[thb]
\begin{center}
\caption{The theoretical predictions (in unit of $10^{-2}$) for the branching ratios of the considered semileptonic decays of $B^0_s$ meson with
$l=(e,\mu)$  obtained by using various theoretical approaches \cite{pqcd2,iamf,rqm,lcsr,lfqm,cqm,qcdsr,labs}. }
\label{tab2}
\vspace{0.2cm}
\begin{tabular}{l| c c | c c } \hline \hline
 Approach   & ${\cal B}(B^0_s \to D^-_s l^+ \nu_l)$ & ${\cal B}(B^0_s \to D^-_s \tau^+ \nu_\tau)$
& ${\cal B}(B^0_s \to D^{*-}_s l^+ \nu_l)$ & ${\cal B}(B^0_s \to D^{*-}_s \tau^+ \nu_\tau )$  \\
\hline
PQCD               & $1.97^{+0.89}_{-0.66}$ & $0.72^{+0.32}_{-0.23}$ & $5.04^{+1.62}_{-1.43}$ & $1.45^{+0.46}_{-0.40}$  \\
PQCD+Lattice       & $1.84^{+0.77}_{-0.51}$ & $0.63^{+0.17}_{-0.13}$ & $4.42^{+1.27}_{-1.00}$ & $1.20^{+0.26}_{-0.23}$ \\
\hline
PQCD\cite{pqcd2}   & $2.13^{+1.12}_{-0.77}$ & $0.84^{+0.38}_{-0.28}$ & $4.76^{+1.87}_{-1.49}$ & $1.44^{+0.51}_{-0.42}$ \\
IAMF\cite{iamf}    & $1.4-1.7$              & $0.47-0.55$            & $5.1-5.8$              & $1.2-1.3$ \\
RQM\cite{rqm}      & $2.1\pm0.2$            & $0.62\pm0.05$          & $5.3\pm0.5$            & $1.3\pm0.1$ \\
LCSR\cite{lcsr}    & $1.0{+0.4}_{-0.3}$     & $0.33^{+0.14}_{-0.11}$ & $ - $                  & $ - $ \\
LFQM\cite{lfqm}    & $ - $                  & $ - $                  & $5.2\pm0.6$            & $1.3^{+0.2}_{-0.1}$ \\
CQM\cite{cqm}      & $2.73-3.00$            & $ - $                  & $7.49-7.66$            & $ - $ \\
QCDSR\cite{qcdsr}  & $2.8-3.8$              & $ - $                  & $1.89-6.61$            & $ - $ \\
Lattice\cite{labs} & $2.013-2.469$          & $0.619-0.724$          & $ - $                  & $ - $ \\
\hline\hline
\end{tabular}
\end{center} \end{table}
\begin{table}[thb]
\begin{center}
\caption{The theoretical predictions for the ratios $R(D_s)$ and $R(D_s^*)$ obtained by using the various theoretical
approaches \cite{pqcd2,rqm,lcsr,lfqm,labs}. }
\label{tab3}
\vspace{0.2cm}
\begin{tabular}{l| c c |c c c c c } \hline \hline
 Ratios        & PQCD &PQCD+Lattice & PQCD\cite{pqcd2} &  RQM\cite{rqm}  & LCSR\cite{lcsr}  & LFQM\cite{lfqm} & Lattice\cite{labs}  \\ \hline
$R(D_s)$   & $ 0.365^{+0.009}_{-0.012} $ & $0.341^{+0.024}_{-0.025}$ & $0.392(22)$ & $0.295$ & $0.33$ & $ - $ & $0.299^{+0.027}_{-0.022}$  \\
$R(D^*_s)$ & $ 0.287^{+0.008}_{-0.011} $ & $0.271^{+0.015}_{-0.016}$ & $0.302(11)$ & $0.245$ & $ - $ & $0.25$ & $-$  \\
\hline\hline
\end{tabular}
\end{center} \end{table}

\begin{table}[thb]
\begin{center}
\caption{The PQCD and ``PQCD+Lattice" predictions for the branching ratios (in unit of $10^{-2}$ )  of the eight semileptonic decays
 $B \to D^{(*)} \tau^+  \nu_\tau$ and  $B \to D^{(*)} l^+ \nu_l$  with $l=(e,\mu)$. As a comparison, we also
 list the previous ``PQCD+Lattice" predictions\cite{pqcd3},  the SM predictions based on HQET \cite{np1} ,
 and the  world  average of the measured values as given in PDG 2018 \cite{pdg2018}.  }
\label{tab4}
\vspace{0.2cm}
\begin{tabular}{l| c c| c c |c} \hline \hline
    Channels           &  PQCD &  PQCD+Lattice & PQCD\cite{pqcd3} &                                    HQET\cite{np1}  & PDG\cite{pdg2018} \\ \hline
$B^+ \to D^0 \tau^+ \nu_\tau$ & $0.86^{+0.34}_{-0.25}$ & $0.69^{+0.21}_{-0.17}$ & $0.95^{+0.37}_{-0.31}$ & $ 0.66\pm 0.05 $ & $0.77\pm0.25$ \\
$B^+ \to D^0 l^+ \nu_l$ & $ 2.29^{+0.91}_{-0.72}$ & $2.10^{+0.85}_{-0.62}$ & $2.19^{+0.99}_{-0.76}$ & $-$ & $2.20\pm0.10$ \\
\hline
$B^+ \to D^{*0} \tau^+ \nu_\tau $ & $1.60^{+0.39}_{-0.37}$ & $1.34^{+0.26}_{-0.23}$ & $1.47^{+0.43}_{-0.40}$ & $1.43\pm 0.05$ & $1.88\pm0.20$ \\
$B^+ \to D^{*0} l^+ \nu_l$ & $5.53^{+1.45}_{-1.25}$ & $4.89^{+1.21}_{-1.00}$ & $4.87^{+1.60}_{-1.41}$ & $-$ & $4.88\pm0.10$ \\
\hline
$B^0 \to D^- \tau^+ \nu_\tau$ & $0.82^{+0.33}_{-0.24}$ & $0.62^{+0.19}_{-0.14}$ & $0.87^{+0.34}_{-0.28}$ & $0.64\pm 0.05 $ & $1.03\pm0.22$ \\
$B^0 \to D^- l^+ \nu_l$ & $2.19^{+0.91}_{-0.68}$ & $1.95^{+0.77}_{-0.56}$ & $2.03^{+0.92}_{-0.70}$ & $-$ & $2.20\pm0.10$ \\
\hline
$B^0 \to D^{*-} \tau^+ \nu_\tau $ & $1.53^{+0.37}_{-0.35}$ & $1.25^{+0.25}_{-0.21}$ & $1.36^{+0.38}_{-0.37}$ & $1.29\pm 0.06 $ & $1.67\pm0.13$ \\
$B^0 \to D^{*-} l^+ \nu_l$ & $5.32^{+1.37}_{-1.20}$ & $4.63^{+1.15}_{-0.95}$ & $4.52^{+1.44}_{-1.31}$ & $-$ & $4.88\pm0.10$ \\
\hline\hline
\end{tabular} \end{center} \end{table}
\begin{table}[thb]
\begin{center}
\caption{The PQCD and ``PQCD+Lattice " predictions for the ratios $R(D)$ and $R(D^*)$.  As a comparison, we also show
the previous ``PQCD+Lattice" predictions as given in Ref.~\cite{pqcd3}, the average of the SM predictions as given in Ref.~\cite{hflav2019},
several measured values as reported by BaBar, Belle and LHCb Collaborations \cite{babar2012,belle2019b,lhcb2018},
as well as the world averaged results from HFLAV  group \cite{hflav2019} }
\label{tab5}
\vspace{0.2cm}
\begin{tabular}{l| c c| c c|c c c c} \hline \hline
Ratios     &  PQCD &  PQCD+Lattice & PQCD\cite{pqcd3} & SM\cite{hflav2019}& BaBar\cite{babar2012}
& Belle\cite{belle2019b} & LHCb\cite{lhcb2018} & HFLAV\cite{hflav2019}  \\ \hline
$R(D)$      & $ 0.376^{+0.011}_{-0.012} $ & $0.324^{+0.020}_{-0.022}$& $0.337(38)$ &$0.299(3)$& $0.440(72)$ & $0.307(40)$ & $ - $                 &$0.340(30)$ \\
$R(D^*)$ & $ 0.288^{+0.008}_{-0.010} $ & $0.272^{+0.013}_{-0.014}$ & $0.269(21)$ &$0.258(5)$& $0.332(30)$ & $0.283(23)$ & $0.291(35)$ & $0.295(14)$ \\
\hline \hline
\end{tabular} \end{center} \end{table}

Following the same procedure as  we did for $B_s$ decays, we can calculate the branching ratios and the ratios   $R(D^{(*)})$
for $B \to (D,D^*)( l^+ \nu_l, \tau^+ \nu_\tau)$ decays by employing the PQCD and the ``PQCD+Lattice" approaches, respectively.
The numerical results are listed in Table  \ref{tab4} for branching ratios and in Table \ref{tab5} for ratios $R(D)$ and $R(D^*)$.
In the estimation for errors,   the parameter  $\omega_{\rm B}=0.40 \pm 0.04$ GeV is used.

From the numerical results as listed in Tables \ref{tab2} -- \ref{tab5}, one can see the following points:
\begin{enumerate}
\item[(1)]
For all considered semileptonic decays $B\to D^{(*)}l^+ \nu_l$ with $l=(e,\mu,\tau)$,  the  ``PQCD+Lattice"  predictions for their branching ratios and the ratios
$R(D^{(*)})$   agree well with those currently available  experimental measurements within the errors, which can be treated as  one evidence for the
reliability of  this ``PQCD+Lattice"  method to deal with the semileptonic decays of $B$ meson.

\item[(2)]
For the four $B_s$ decay modes and the eight $B$ decay modes,   the  ``PQCD+Lattice "  predictions for the branching ratios  are
generally smaller than the conventional PQCD predictions, but the differences between them are relatively small:  less than $20\%$ in magnitude.
These predictions also agree well with previous PQCD predictions as given in Ref.~\cite{pqcd1,pqcd2,pqcd3} within the still large
theoretical uncertainties.

\item[(3)]
For the ratios $R(D_s)$ and $R(D_s^*)$,  the theoretical errors in branching ratios are largely cancelled,
the PQCD and "PQCD+Lattice"  predictions for both  ratios $R(D_s)$ and $R(D_s^*)$ have a  very small error only:
around $5\%$ in magnitude.   These predictions  could be tested in the near future LHCb and Belle-II experiments.

\item[(4)]
For the branching ratios, the theoretical predictions from different theoretical approaches can be a little different for the same decay mode,
but they still agree within errors due to the large theoretical uncertainties.
For the  ratios $R(D)$ and $R(D^*)$, our ``PQCD+Lattice" predictions also agree well with the average of the SM predictions obtained by employing
the HQET plus  the  available Lattice QCD input for the relevant form factors  \cite{sm4,sm1,sm2,sm3,hflav2019}.

\item[(5)]
By comparing  the values of the  two sets of the ratios  $(R(D), R(D_s))$ and $( R(D^*),R(D_s^*) )$,  as listed  in Table \ref{tab3} and \ref{tab5},
one can see that  the SU(3) flavor symmetry  keeps very well  in both the PQCD and the ``PQCD+Lattaice"  approaches.
This point could also be tested  by experiments.

\end{enumerate}

\subsection{$P_\tau(D_{(s)}^{(*)})$,  $F_L(D_{(s)}^*)$ and $A_{FB}(\tau)$}

Besides the branching ratios and the ratios $R(X)$ of the branching ratios,  there are other three additional physical observables:
such as the longitudinal polarization of the tau lepton $P_\tau(D_{(s)}^{(*)})$,  the fraction of $D_{(s)}^*$ longitudinal polarization  $F_L(D_{(s)}^*)$
and the forward-backward asymmetry of the tau lepton $A_{FB}(\tau)$. These physical quantities can be measured in the LHCb and Belle experiments, and
may be sensitive to some kinds of new physics \cite{prd072012,prd114022,prd036021,ptau1,ptau2}.
The calculations and investigations for these additional physical observables may provide new clues to understand the $R(D^{(*)})$ puzzle, and it  is therefore
necessary and interesting.

In Refs.~\cite{ptau2,huang18,bhatt18},  the authors calculated these three physical obsrvables in the framework of the SM and examined possible new physics
effects on them.  In this paper, we calculate  these observables by using the PQCD and the   ``PQCD+Lattice'' approach, respectively.
Based on the formulae as given explicitly in Eqs.~(\ref{eq:ptau}-\ref{eq:btheta2}),
we make the calculations and show the numerical predictions for $P_\tau(D_{(s)}^{(*)})$,   $F_L(D_{(s)}^*)$
and $A_{FB}(\tau)$  in Table~\ref{tab6}.
The measured values of $P_{\tau}(D^*)$ and $F_L(D^*)$ \cite{prl118-801,prd97-012004,1903tau}
as given in Eqs.~(\ref{eq:ptaustar},\ref{eq:flstar}) are also listed in Table \ref{tab6}.  As a comparison,  we also show other
SM predictions for  these physical observables   as given in Refs.~\cite{ptau2,huang18} in Table \ref{tab6}.

\begin{table}[thb]
\begin{center}
\caption{The PQCD and ``PQCD+Lattice"  predictions for   $P_\tau(D_{(s)}^{(*)})$,   $F_L(D_{(s)}^*)$
and $A_{FB}(\tau)$,   other SM predictions and  the measured values are all listed in this table. }
\label{tab6}
\vspace{0.2cm}
\begin{tabular}{l| l| c c|cc  } \hline \hline
Observable      & Approach   &  $B^0\to D^- \tau^+\nu_\tau$ & $B^0_s\to D^-_s \tau^+ \nu_\tau $  & $B^0\to D^{*- } \tau^+ \nu_\tau $
& $B^0_s\to D_s^{*-} \tau^+ \nu_l $  \\ \hline
&PQCD&  $0.32(1)$   &$0.31(1)$& $-0.54(1)$& $-0.54(1)$\\
&PQCD+Lat.& $0.30(1)$& $0.30(1)$& $-0.53(1)$& $-0.53(1)$\\
$P_\tau(D_{(s)}^{(*)})$&SM\cite{ptau2}&$-$&$-$&$-0.497(13)$&$-$\\
&SM\cite{huang18}&$0.325(3)$&$-$&$-0.508(4)$& $-$\\
&Belle\cite{prl118-801}&$-$&$-$&$-0.38\pm 0.51^{+0.21}_{-0.16}$&$-$\\   \hline
&PQCD& $-$&$-$&$0.42(1)$& $0.42(1)$\\
&PQCD+Lat.& $-$&$-$&$0.43(1)$& $0.43(1)$\\
$F_{\rm L}(D_{(s)}^*)$&SM\cite{bhatt18}& $-$&$-$&$0.457(10)$&$-$\\
&SM\cite{huang18}&$-$&$-$&$0.441(6)$&$-$\\
&Belle\cite{1903tau}&$-$&$-$&$0.60\pm 0.08\pm 0.04$&$-$\\   \hline
                                         &PQCD&$0.35(1)$&$0.36(1)$&$-0.085(2)$&$-0.083(2)$\\
$A_{\rm FB}(\tau) $ &PQCD+Lat.&$0.36(1)$&$0.36(1)$&$-0.054(2)$&$-0.050(2)$\\
&SM\cite{huang18}&$0.361(1)$&$-$&$-0.084(13)$&$-$\\
\hline\hline
\end{tabular}
\end{center} \end{table}

From the numerical results as listed in  Eqs.~(\ref{eq:ptaustar},\ref{eq:flstar}) and in Table~\ref{tab6},  one can find the following points:
\begin{itemize}
\item[(1)]
The  uncertainties of  all theoretical predictions for  the physics observables $P_\tau(D_{(s)}^{(*)})$, $F_L(D_{(s)}^*)$ and $A_{FB}(\tau)$
are very small when compared with the ones for the branching ratios,  since the theoretical uncertainties are largely cancelled in ratios.

\item[(2)]
The  PQCD  and   ``PQCD+Lattice''  predictions for  the physics observables $P_\tau(D_{(s)}^{(*)})$, $F_L(D_{(s)}^*)$ and $A_{FB}(\tau)$
for $B_{(s)} \to D_{(s)}\tau^+ \nu_\tau$ decays are very similar with each other: the difference for a fixed decay mode is less than $5\%$.
For  the $A_{FB}(\tau)$  of  the $B_{(s)} \to D^*_{(s)} \tau^+\nu_\tau$ decays, however,  the difference is about $40\%$.
 The reason  lies in the definition of the angular coefficient function $b^{(D^*)}_\theta(q^2)$,
where the term $(H_{V,+}^2-H_{V,-}^2)$ in Eq.~(\ref{eq:btheta2})  can  be affected moderately by the different high $q^2$ behaviour of the form factors
in the PQCD and ``PQCD+Lattice" approach.  Of course, the precise experimental measurement of  the forward-backward  asymmetry
$A_{FB}(\tau)$  of  the $B_{(s)} \to D^*_{(s)} l^+\nu_l$ decays will be a great help for us to test and improve the factorization models. 

\item[(3)]
For both  $P_\tau(D^*)$ and $F_L(D^*)$, our  theoretical predictions agree well with the measured ones within errors,
partially due to the still  large experimental errors.
For all considered decay modes, the PQCD and ``PQCD+Lattice" predictions  for the three kinds of physics observables are consistent with the results from other approaches in the framework of the SM as well.
We expect that these physical observables could be measured in high precision at the future LHCb and Belle-II experiments
and it can help us to test the theoretical models or approaches.

\end{itemize}

\section{Summary} \label{sec:4}

In this paper, we studied the semileptonic decays $B_{(s)} \to D_{(s)}^{(*)} l^+ \nu_l $ in the frame work of the SM
by employing both the conventional PQCD factorization approach and the ``PQCD+Lattice'' approach.
In the second approach, we take into account the Lattice QCD results of  the relevant form factors as an input in the extrapolation  from
the low $q^2$ to the  endpoint $q^2_{max}$.
We calculated the form factors $F_{\rm 0,+}(q^2)$, $V(q^2)$ and $A_{\rm 0,1,2}(q^2)$ of  the
$B_{(s)} \to D_{(s)}^{(*)} $ transitions,  provided the  theoretical predictions for the  branching ratios of the considered $B/B_s$ semileptonic decays
and the ratios $R(D^{(*)})$ and $R(D_s^{(*)})$.
In addition to the branching ratios and the ratios $R(X)$, we also gave our theoretical predictions for the additional physical observables:
the longitudinal polarization  of the tau lepton $P_\tau(D_{(s)}^{(*)})$,  the fraction of $D_{(s)}^*$ longitudinal polarization  $F_L(D_{(s)}^*)$
and the forward-backward asymmetry of the tau lepton $A_{FB}(\tau)$.

From the numerical calculations and phenomenological analysis we found the following points:
\begin{itemize}
\item[(1)]
For the twelve considered $B/B_s$ semileptonic decay modes,   the  ``PQCD+Lattice "  predictions for the branching ratios  are
generally smaller than the conventional PQCD predictions, but the differences between them are relatively small:  less than $20\%$ in magnitude.
The  ``PQCD+Lattice"  predictions for their branching ratios and the ratios $R(D^{(*)})$  do  agree well with those currently available
experimental measurements within the errors.

\item[(2)]
For the ratios $R(D_s)$ and $R(D_s^*)$,  the PQCD and "PQCD+Lattice"  predictions are the following:
\beq
R(D_s)&=& \left \{ \begin{array}{ll} 0.365^{+0.009}_{-0.012} ,  &  {\rm PQCD}, \\    0.341^{+0.024}_{-0.025} ,   & {\rm PQCD+Lattice},\\
 \end{array} \right.  , \label{eq:rds1a}  \\
R(D^*_s)&=&\left \{ \begin{array}{ll}   0.287^{+0.008}_{-0.011} ,  &  {\rm PQCD}, \\    0.271^{+0.015}_{-0.016} ,   & {\rm PQCD+Lattice},\\
 \end{array} \right.  . \label{eq:rds1b}
\eeq
They also agree well with other SM predictions based on different approaches.  These predictions  could be tested by the forthcoming
LHCb and Belle-II experiments.

\item[(3)]
For  most observables $P_\tau(D_{(s)}^{(*)})$, $F_L(D_{(s)}^*)$ and $A_{FB}(\tau)$, the  PQCD  and   ``PQCD+Lattice''  predictions
agree very well with each other: the difference is less than $5\%$.  The relatively large difference for  the $A_{FB}(\tau)$
of the $B_{(s)} \to D^*_{(s)} \tau^+\nu_\tau$ decays can be understood  reasonably.

\item[(4)]
For both  $P_\tau(D^*)$ and $F_L(D^*)$, our  theoretical predictions agree well with the measured ones within errors.
For all considered decay modes, the PQCD and ``PQCD+Lattice" predictions  for the three kinds of physics observables are consistent
with the results from other approaches in the framework of the SM as well.
The future LHCb and Belle-II experiments can help us to test the above theoretical predictions.

\end{itemize}

\begin{acknowledgments}
We wish to thank Wen-Fei Wang and Ying-Ying Fan for valuable discussions. This work was supported by the National Natural Science Foundation of China under Grant  No.~11775117 and 11235005.
\end{acknowledgments}


\appendix
\section{Relevant functions}\label{sec:app1}
In this appendix, we present the explicit expressions for some functions which have already appeared in the previous sections.
The hard functions $h_1$ and $h_2$ come form the Fourier transform, and they can be written as:
\beq
\begin{aligned}
h_1(x_1,x_2,b_1,b_2)&=K_0(\beta_1 b_1)
[\theta(b_1-b_2)I_0(\alpha_1 b_2)K_0(\alpha_1 b_1)\\
&+\theta(b_2-b_1)I_0(\alpha_1 b_1)K_0(\alpha_1 b_2)]S_t(x_2),
\end{aligned}
\eeq
\beq
\begin{aligned}
h_2(x_1,x_2,b_1,b_2)&=K_0(\beta_2 b_1)
[\theta(b_1-b_2)I_0(\alpha_2 b_2)K_0(\alpha_2 b_1)\\
&+\theta(b_2-b_1)I_0(\alpha_2 b_1)K_0(\alpha_2 b_2)]S_t(x_2),
\end{aligned}
\eeq
where $K_0$ and $I_0$ are modified Bessel functions, and
\beq
\alpha_1 = m_{B_{(s)}}\sqrt{x_2r \eta^+},\quad
\alpha_2=m_{B_{(s)}}\sqrt{x_1 r \eta^+ - r^2+r_c^2},\quad
\beta_1 = \beta_2=m_{B_{(s)}}\sqrt{x_1x_2 r \eta^+}.
\eeq

The threshold resummation factor $S_t(x)$ is adopted from Ref.~\cite{prd65-014007}:
\beq
S_t=\frac{2^{1+2c}\Gamma(3/2+c)}{\sqrt{\pi}\Gamma(1+c)}[x(1-x)]^c,
\eeq
and we here set the  parameter $c=0.3$.

The factor $\exp[-S_{ab}(t)]$ contains the Sudakov logarithmic
corrections and the renormalization group evolution effects of both
the wave functions and the hard scattering amplitude with
$S_{ab}(t)=S_B(t)+S_M(t)$ \cite{prd65-014007,prd63-074009},
\beq
S_B(t)&=&s\left(x_1\frac{m_{B_{(s)}}}{\sqrt{2}},b_1\right)
+\frac{5}{3}\int_{1/b_1}^{t}\frac{d\bar{\mu}}{\bar{\mu}}\gamma_q(\alpha_s(\bar{\mu})),\\
S_M(t)&=&s\left (x_2\frac{m_{B_{(s)}}}{\sqrt{2}} r\eta^+,b_2 \right )
+s\left ((1-x_2)\frac{m_{B_{(s)}}}{\sqrt{2}} r \eta^+,b_2 \right)
+2\int_{1/b_2}^{t}\frac{d\bar{\mu}}{\bar{\mu}}
\gamma_q(\alpha_s(\bar{\mu})),
\eeq
where $\eta^+$ is defined in Eq.~(\ref{eq2}).
The hard scale $t$ and the quark anomalous dimension $\gamma_q=-\alpha_s/\pi$ governs the aforementioned renormalization group evolution.
The explicit expressions of the functions $s(Q,b)$ can be found for example in Appendix A of
Ref.~\cite{prd63-074009}. The hard scales $t_i$ in Eqs.(\ref{eqf1}-\ref{eqA2}) are chosen as the largest scale of
the virtuality of the internal particles in the hard $b$-quark decay diagram,
\beq
t_1&=&\max\{m_{B_{(s)}}\sqrt{x_2 r \eta^+}, 1/b_1, 1/b_2\},\non
t_2&=&\max\{m_{B_{(s)}}\sqrt{x_1 r \eta^+ - r^2+r_c^2},1/b_1, 1/b_2\}.
\eeq


\end{document}